\documentclass[prl,aps,reprint,footnotebib,amssymb]{revtex4-2}
\usepackage{amsmath}
\usepackage{bm}
\usepackage[dvipsnames]{xcolor}
\usepackage{graphicx}
\usepackage{nicematrix}
\usepackage{epstopdf}
\usepackage{epsfig}
\usepackage{amsfonts}
\usepackage[pdfencoding=auto,naturalnames]{hyperref}
\usepackage{hypcap}
\usepackage{mathtools}
\usepackage{verbatim}
\usepackage{tabularx}
\usepackage{bbm}
\usepackage{esvect}
\usepackage{subfigure}
\usepackage{slashed}
\usepackage{physics}
\usepackage{times}
\usepackage{multirow}
\usepackage{float}
\usepackage{cancel}

\hypersetup{colorlinks=true, citecolor=Blue, urlcolor=Blue, linkcolor=Blue, breaklinks=true}

\newcommand{\ssll}{\hspace{-1em}---}

\def\bea{\begin{eqnarray}}
\def\eea{\end{eqnarray}}
\def\be{\begin{equation}}
\def\ee{\end{equation}}
\def\nn{\nonumber}
\def\ba{\begin{array}}
\def\ea{\end{array}}
\def\Tr{\mathrm{Tr}}
\def\sgn{\text{sgn}}

\begin{document}
\title{Boundary criticality for the Gross-Neveu-Yukawa models}

\author{Huan Jiang}
\author{Yang Ge}
\author{Shao-Kai Jian}
\email{sjian@tulane.edu}
\affiliation{Department of Physics and Engineering Physics, Tulane University, New Orleans, Louisiana 70118, USA}

\date{\today}

\begin{abstract}
    We study the boundary criticality for the Gross-Neveu-Yukawa (GNY) models. 
    Employing interacting Dirac fermions on a honeycomb lattice with armchair boundaries, we use determinant quantum Monte Carlo simulation to uncover rich boundary criticalities at the quantum phase transition to a charge density wave (CDW) insulator, including the ordinary, special, and extraordinary transitions. 
    The Dirac fermions satisfy a Dirichlet boundary condition, while the boson field, representing the CDW order, obeys Dirichlet and Neumann conditions at the ordinary and special transitions, respectively, thereby enriching the critical GNY model. 
    We develop a perturbative $4-\epsilon$ renormalization group approach to compute the boundary critical exponents. 
    Our framework generalizes to other GNY universality class variants and provides theoretical predictions for experiments. 
\end{abstract}

\maketitle

\paragraph{Introduction}\ssll
Boundary criticality captures the rich phenomenology of boundary effects in conformal field theory (CFT)~\cite{andrei2018boundarydefectcftopen}. 
It is well known that the presence of boundaries enriches universality classes. 
For instance, in a free scalar field theory, there exist two distinct conformal boundary conditions, Dirichlet and Neumann, each leading to different boundary scaling dimensions for the scalar field. 
The situation becomes more intriguing in interacting theories, such as the interacting scalar CFT with $O(N)$ symmetry~\cite{diehl1981field,diehl1983multicritical,domb1986phase,diehl1996the,cardy1996scaling,liendo2012the,giombi2020cft}, in which a much richer landscape of boundary critical phenomena arises, broadly categorized under boundary CFT (BCFT). 
BCFT has a wide range of applications across various fields of physics, from the Kondo effect in condensed matter systems~\cite{kondo1964resistance,affleck1991the} to holographic dualities in quantum gravity~\cite{maldacena1997the,witten1998anti,gubser1998gauge}. 
Consequently, exploring new boundary criticalities holds great significance across multiple disciplines.

Gapless Dirac fermions, which emerge in numerous condensed matter systems, including graphene~\cite{castro2009the}, the surface of topological insulators~\cite{hasan2010colloquium,qi2011topological}, and topological semimetals~\cite{armitage2018weyl,lv2021experimental}, can drive exotic quantum phase transitions. 
For instance, the Ising transition is enriched to the chiral Ising universality class in the presence of Dirac fermions~\cite{rosenstein1993critical,herbut2024wilson}. 
In these critical theories, the scalar field couples to Dirac fermions via a Yukawa interaction, providing a realization of the Gross-Neveu-Yukawa (GNY) model~\cite{gross1974dynamical,zinn-Justin1991four,moshe2003quantum, herbut2009relativistic}.
It has generated extensive research interest in the context of BCFT~\cite{mcavity1993energy,carmi2018a,herzog2017boundary,herzog2018superconformal,schaub2023spinors,barrat2023line,kakkar2023phases,brillaux2023surface,shen2024new}. 
In particular, previous works~\cite{giombi2021fermions,herzog2023fermions} pioneered the study of boundary criticality in the GNY universality class, identifying distinct boundary transitions, including special, ordinary, and extraordinary boundary conditions, when Dirac fermions couple to a scalar field. 
However, it remains unclear how this theory describes boundary criticality in condensed matter systems and other variants of the GNY universality class. 

In this paper, we systematically address these questions. 
We construct a lattice model that hosts interacting Dirac fermions and an Ising transition with an open boundary condition. 
Using determinant quantum Monte Carlo (DQMC) simulations~\cite{assaad2020ALF}, we demonstrate that the model exhibits rich boundary criticalities, including the ordinary, special, and extraordinary transitions. 
These transitions are identified by examining renormalization-group-(RG) invariant quantities in the bulk and at the boundary. 
This, to the best of our knowledge, represents the first realization of boundary criticality of the GNY model on a lattice model. 
To formulate an effective field theory, we identify that the Dirac fermion satisfies a Dirichlet boundary condition~\footnote{Note that the boundary condition of Dirac fermions has been extensively discussed in condensed matter physics~\cite{brey2006eletronic,akhmerov2008boundary,faraei2018green,shtanko2018robustness}.}, while the scalar field obeys Dirichlet and Neumann boundary conditions at the ordinary and special transitions, respectively. 
These conformal boundary conditions enrich the critical GNY model.
In particular, we develop a perturbative $4-\epsilon$ RG approach to compute the critical exponents. 
Our method is broadly applicable to other variants of the GNY universality class, including the chiral XY and chiral Heisenberg universality classes. 
Notably, our critical theory differs from previous works~\cite{giombi2021fermions,herzog2023fermions} due to the symmetry of the order parameters.
As a result, we present, for the first time, the boundary critical exponents of the GNY universality class in the context of condensed matter theory, as summarized in Table~\ref{tab:scaling}.

\begin{table*}
	\begin{tabular}{c|c|c|c|c|c}
		\hline
		\hline
		\multirow{2}{*}{GNY}  & \multicolumn{2}{c|}{Ordinary transition} & \multicolumn{3}{c}{Special transition} \\
		\cline{2-6}
		&  $\Delta_{\hat \psi} $ & $\Delta_{ \partial \hat\phi}$ &  $\Delta_{\hat \psi} $ & $\Delta_{\hat \phi}$ & $\Delta_{\hat \phi^2}$\\
		\hline
		\hline
		CI &$\frac23-\frac{5+4N}{12+8N}\epsilon$ &$2 -\frac{21+22N+\sqrt{4N^2+132N+9}}{12(3+2N)}\epsilon$ & $\frac32-\frac{3+4N}{12+8N}\epsilon $& $1-\frac{21-2N+\sqrt{4N^2+132N+9}}{12(3+2N)}\epsilon$ & $2-\frac{15-4N-\sqrt{4N^2+312N+9}}{6(3+2N)}\epsilon$  \\
		\hline
		CH & $\frac32-\frac{8N-1}{4+16N}\epsilon$ & $2-\frac{3(9+52N)+5\sqrt{16N^2+344N+1}}{44(1+4N)}\epsilon$ & $\frac32-\frac{8N-7}{4+16N}\epsilon$ & $1-\frac{27-20N+5\sqrt{16N^2+344N+1}}{44(1+4N)}\epsilon$ & $2-\frac{17-24N-5\sqrt{16N^2+344N+1}}{22(1+4N)}\epsilon$ \\
		\hline
	\end{tabular}
	\caption{Boundary scaling dimensions for the GNY universality class, including the chiral Ising (CI) and the chiral Heisenberg (CH) universality class.
		$\epsilon = 4 -d$, and $N$ denotes the flavor of four-component fermions in CI and eight-component fermions in CH. 
	}
	\label{tab:scaling}
\end{table*}

\begin{figure}[t]
    \centering
    \subfigure[]{
    \includegraphics[width=0.52\linewidth]
    {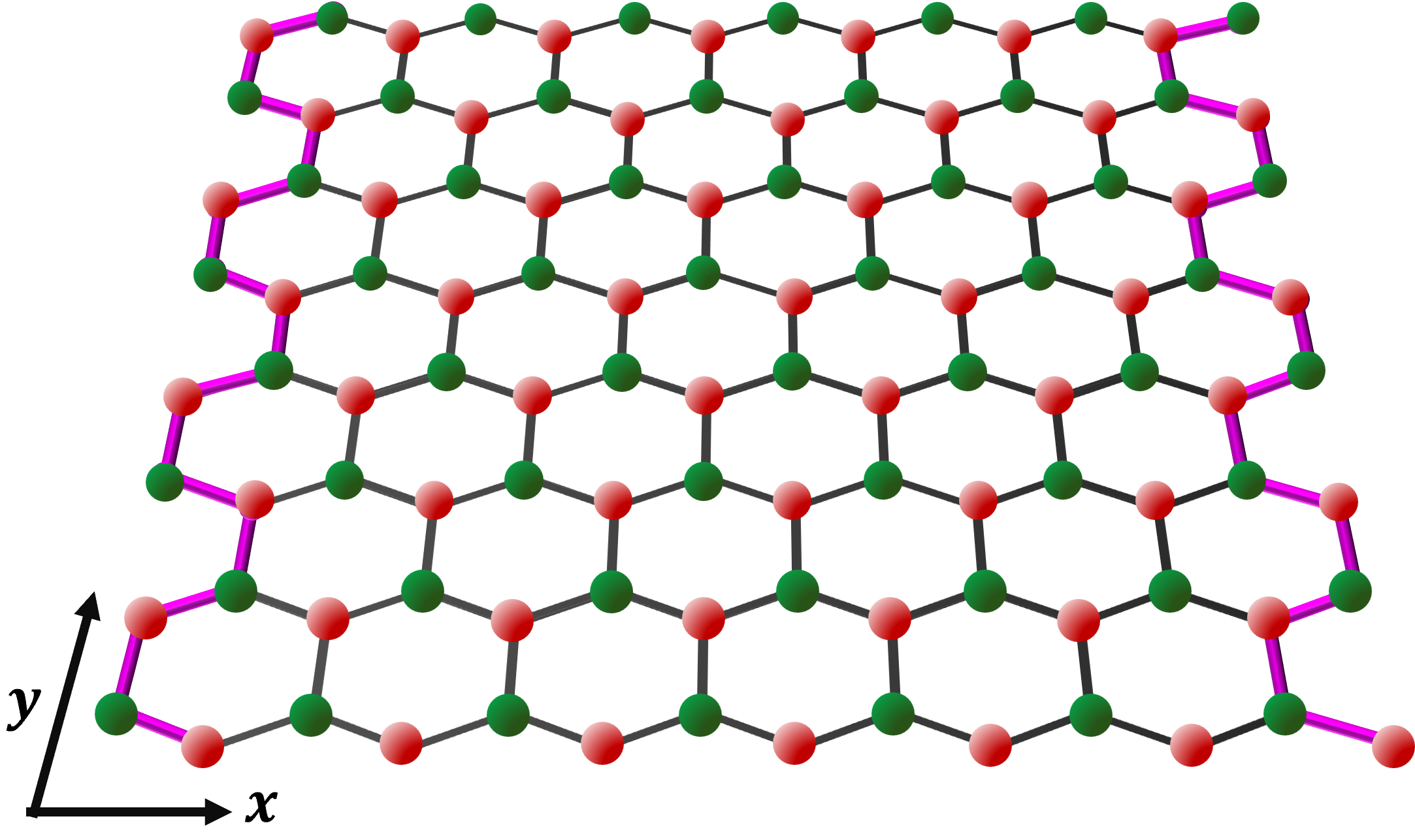}
    } \\
    \subfigure[]{\includegraphics[width=0.5\linewidth]
    {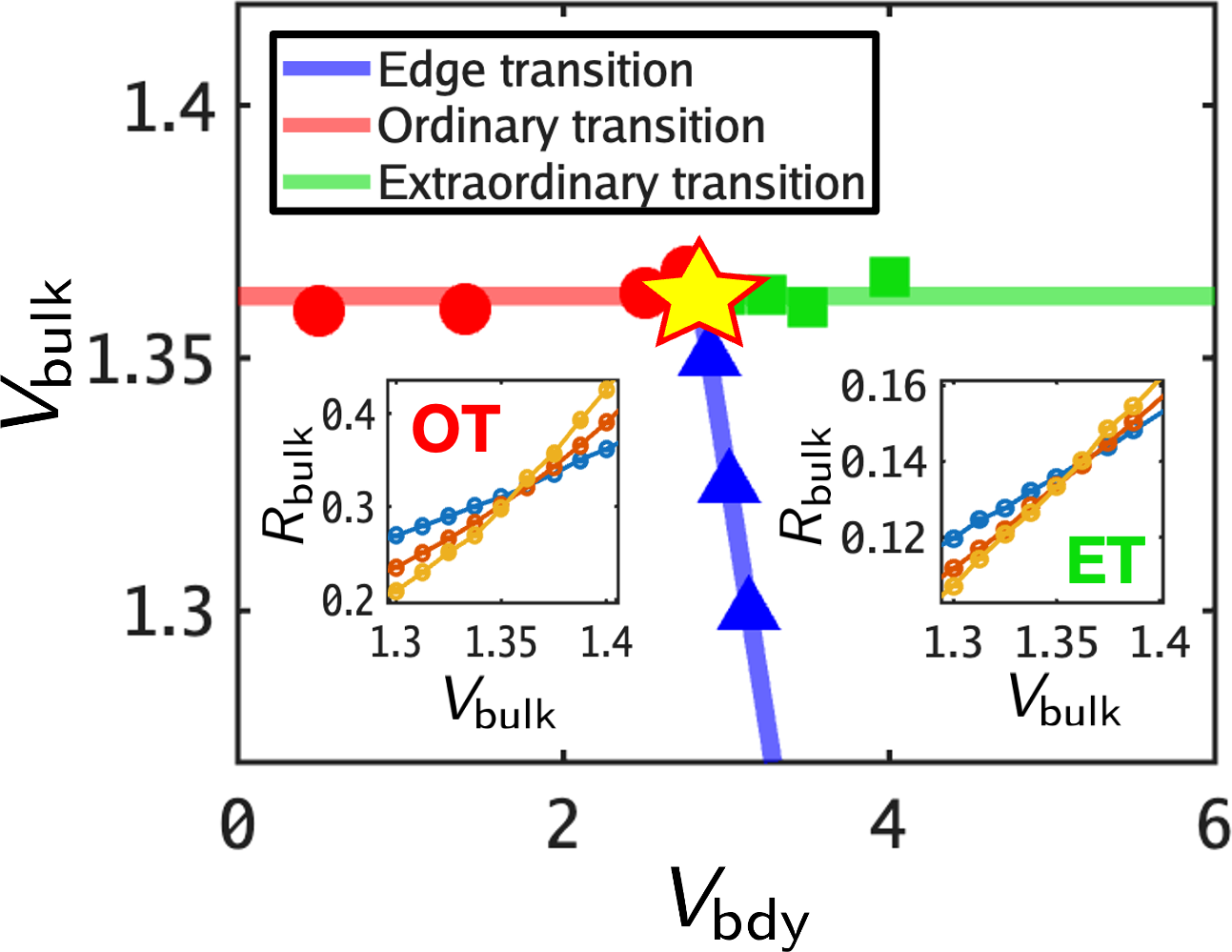}} \qquad
    \subfigure[]{\includegraphics[width=0.408\linewidth]
    {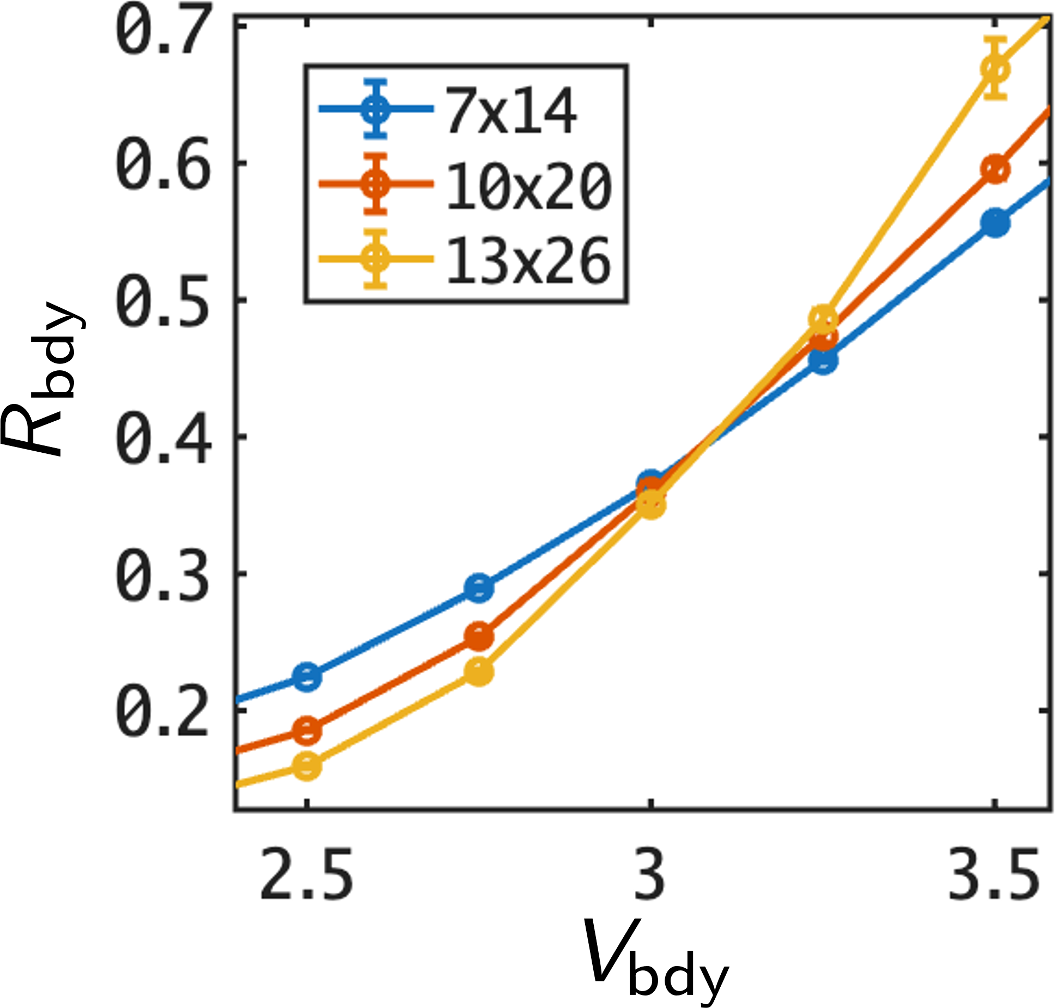}}
    \caption{(a) Illustration of the honeycomb lattice ($L_x=7$) with armchair boundaries. The black (pink) link denotes the bulk (boundary) bond. The red and green sites denote the two sublattices, respectively. (b) Phase diagram of the lattice model~\eqref{eq:tV-model}. 
    The inset illustrates the crossing of the RG-invariant quantity $R_\text{bulk}$, where OT (ET) denotes the ordinary transition (extraordinary transition). 
    (c) The crossing of the boundary RG-invariant quantity $R_\text{bdy}$ at the edge transition, with $V_\text{bulk}=1.3$ fixed.}
    \label{fig:lattice-model}
\end{figure}

\paragraph{Lattice model}\ssll
We consider a system of spinless fermions on a honeycomb lattice with armchair boundaries described by the Hamiltonian:
\begin{align}
	\label{eq:tV-model}
    H =& - t \sum_{\langle \boldsymbol{ij} \rangle } ( c_{\boldsymbol i}^\dag c_{\boldsymbol j} + \text{H.c.}) + V_\text{bulk} \sideset{}{'}\sum_{\langle \boldsymbol{ij} \rangle } \left(n_{\boldsymbol i} - \frac12 \right)\! \left(n_{\boldsymbol j} - \frac12 \right) \nn  \\ 
    & + V_\text{bdy} \sideset{}{''}\sum_{\langle \boldsymbol{ij} \rangle } \left(n_{\boldsymbol i} - \frac12 \right)\! \left(n_{\boldsymbol j} - \frac12 \right),
\end{align}
where $c_{\boldsymbol i}^\dag$ ($c_{\boldsymbol i}$) creates (annihilates) a fermion at site $\boldsymbol i$, $\langle \boldsymbol{ij} \rangle$ denotes a nearest neighbor bond connecting site $\boldsymbol i$ and $\boldsymbol j$, and $n_{\boldsymbol{i}}=c_{\boldsymbol{i}}^\dagger c_{\boldsymbol{i}}$.
The parameter $t$ represents the hopping amplitude. 
The armchair boundaries of the honeycomb lattice are illustrated in Fig.~\ref{fig:lattice-model}, where the black and pink links represent the bulk and boundary bonds, respectively. 
The parameters $V_\text{bulk}$ and $V_\text{bdy}$ are the interaction strengths at bulk and boundary bonds, respectively, with the summation $\sum'_{\langle \boldsymbol{ij} \rangle }$ and $\sum''_{\langle \boldsymbol{ij} \rangle }$ covering bulk and boundary bonds accordingly. 

We consider a lattice geometry with $L_y=2L_x$, in which $L_{x,y}$ denotes the number of unit cells along $x$ and $y$, respectively. 
Using DQMC simulation, we obtain the phase diagram shown in Fig.~\ref{fig:lattice-model}(b). 
As the bulk interaction strength $V_\text{bulk}$ increases, the system undergoes a quantum phase transition at $V^\ast_\text{bulk}$ from a Dirac semimetal to a charge density wave (CDW) insulator, characterized by the CDW order parameter 
$\Delta_{\boldsymbol{i}}\equiv (-1)^{s(\boldsymbol{i})}\left( n_{\boldsymbol{i}}-\frac12\right)$, and $s(\boldsymbol{i}) = 0, 1$ for two sublattices. 
The bulk critical point is determined by the crossing of the RG-invariant quantity, defined by
\begin{equation}
\label{eq:rginv}
    R_{\rm w}\equiv \frac{1}{2\pi}\sqrt{|\tilde C_{\rm w}(0)/{\rm Re}\,\tilde{C}_{\rm w}(k_{\rm min})|-1},
\end{equation}
where the minimal momentum $k_{\rm min}=4\pi/L_y$. Note that the unit cell size along $y$ effectively doubles in an armchair. Consequently $k_{\rm min}$ also doubles. 
Let $\boldsymbol r\equiv(x,y)$ and $\boldsymbol r'$ be the displacements to lattice sites, then
\begin{equation}
\label{eq:ck}
\tilde C_{\rm w}(k)=\sum_{\boldsymbol r,\boldsymbol r' \in {\rm w}}\langle\Delta_{\boldsymbol r}\Delta_{\boldsymbol r'}\rangle e^{-ik(y-y')}.
\end{equation}
Here, ${\rm w}\in\{\text{bulk},\text{bdy}\}$ denotes the bulk or boundary regions, respectively. 
As shown in the insets of Fig.~\ref{fig:lattice-model}(b), the DQMC results reveal a clear $R_\text{bulk}$ crossing around $V_\text{bulk}^\ast=1.36$, signaling a bulk quantum phase transition.    %

In addition, we investigate the edge (boundary) transition in the regime before the bulk becomes ordered, i.e., $V_\text{bulk} \le V_\text{bulk}^\ast$. 
The edge transition point is determined by the crossing of the boundary RG-invariant quantity $R_\text{bdy}$, as shown in Fig.~\ref{fig:lattice-model}(c). 
As the bulk interaction increases, the edge transition points gradually shift to lower values and eventually merge with the bulk transition point, marking the special transition, as indicated by the star in Fig.~\ref{fig:lattice-model}(b). 
The DQMC results demonstrate that the edge can order prior to the bulk if the coupling at the edge is big enough. 
Hence, at the bulk transition line $V_\text{bulk}=V_\text{bulk}^*$, the boundary order parameter $\hat\Delta=\sum_{\boldsymbol i\in\text{bdy}}\Delta_{\boldsymbol{i}}$ becomes nonzero in the thermodynamic limit when the boundary interaction strength $V_\text{bdy}$ exceeds a critical value $V_\text{bdy}^*$, where
the special transition occurs. 
For $V_\text{bdy}<V_\text{bdy}^*$, the boundary order remains zero, corresponding to the ordinary transition. 
In contrast, when $V_\text{bdy}>V_\text{bdy}^*$, the boundary order acquires a finite value, marking the extraordinary transition. 
We also carry out a mean field calculation to show the boundary phase diagram in Supplemental Material, which is fully consistent with the DQMC results.

\paragraph{Field theory}\ssll
It is well known that the bulk quantum phase transition falls within the chiral Ising universality class, described by the action:
\begin{align}
	\label{eq:bulk-action}
    S = \int_\mathcal{M}d^dx \Bigg( & \sum_j\bar\psi_j\gamma^\mu\partial_\mu\psi_j 
    + \frac12 (\partial_\mu \phi)^2 + \frac{\lambda}{4!} \phi^4 \nn \\ 
    & -ig\sum_j\phi\bar\psi_j\gamma^5\psi_j \Bigg),
\end{align}
where $\psi_j$ represents the Dirac fermion fields and $\phi$ is the order parameter field~\footnote{We set the velocity to be one due to the emergent Lorentz symmetry. The presence of the boundary will not affect the bulk renormalization.}.
$\partial_\mu$ denotes derivatives regarding (imaginary) time $x_0 \equiv \tau$ and space $x_i$, $i=1,...,d-1$. 
$\gamma_\mu$ are gamma matrices: $\gamma^0=\tau^y\sigma^x$, $\gamma^1=\tau^x$, $\gamma^2=\tau^y\sigma^z$, $\gamma^3=\tau^z$, and $\gamma^5=\tau^y\sigma^y$, and $\bar \psi = \psi^\dag \gamma^0$. 
The Pauli matrices $\sigma^i$ and $\tau^i$ characterize the sublattice and valley degrees of freedom, respectively
This convention of $\gamma$ matrices is introduced in Supplemental Material~\footnote{Our convention of the $\gamma$ matrices is different from, e.g., Ref.~\cite{zerf2017four-loop}, but the results do not depend on the convention of $\gamma$ matrices.}.
Note that, to make our theory more general, we extend the fermion sector to include $N$ flavors, denoted as $\psi_j$ for $j=1,...,N$. 
The parameters $\lambda$ and $g$ represent the quartic boson self-interaction and the fermion-boson Yukawa coupling, respectively.
The Yukawa coupling can be understood as follows: the CDW order corresponds to a staggered density between two sublattices, which is captured by $\sigma^z = - i \gamma^0 \gamma^5$.

We assume that the field theory is located in a $d$-dimensional semi-infinite spacetime, $\mathcal M = \{ x_\mu | x_1>0 \}$ where the boundary is $\partial \mathcal M = \{ x_\mu| x_1=0 \}$. 
For convenience, we denote the hyperspace parallel to the boundary as $y$ and the coordinate perpendicular to the boundary as $x$. 
The boundary condition for the Dirac fermion arises from the vanishing of the lattice fermion outside the lattice sites, enforcing a Dirichlet boundary condition~\cite{brey2006eletronic}, and for completeness, we review its derivation in Supplemental Material. 
In our convention, the boundary condition for the Dirac fermion field is $-\gamma^1 \psi|_\text{bdy} = \psi|_\text{bdy}$. 
For the boson field, the boundary condition can be implemented through a boundary mass term, given by  $\int_{\partial \mathcal M} d^{d-1} x \frac{c}2 \phi^2$, where $c$ is the boundary mass. 
This is connected to the boundary transitions: when $c>0$, the boundary order remains zero corresponding to the ordinary transition, while for $c<0$, the boundary order becomes nonzero corresponding to the extraordinary transition, and $c=0$ corresponds to the special transition~\footnote{Notice that this is a tree-level analysis~\cite{diehl1981field,diehl1983multicritical,domb1986phase}}. 

The effect of these boundary conditions is reflected in the bare propagators for the fermion and boson fields, which are detailed in Supplemental Material. 
We summarize the key results for the bare propagators.
Because of translation symmetry in the hyperspace parallel to the boundary, the propagator can be conveniently expressed in a mixed representation, using momentum $k_\mu$ (for $\mu\ne1$) along the boundary and a real-space coordinate perpendicular to the boundary.
The fermion propagator reads
\begin{eqnarray}
	\label{eq:fermion-prop}
    G(k,x,x')&=&G_b(k,x,x')+G_s(k,x,x'), \\
    G_b(k,x,x') &=& \left(\frac{i\slashed{k}}{2q}-\frac{\gamma^1}{2} \sgn(x-x') \right)e^{-q|x-x'|},\\
    G_s(k,x,x') &=&-\gamma^1 \left(\frac{i\slashed{k}}{2q}+\frac{\gamma^1}{2} \right)e^{-q(x+x')},
\end{eqnarray}
where $\slashed k = \sum_{\mu \ne 1} k_\mu \gamma^\mu$ and $q^2 = \sum_{\mu \ne 1} k_\mu^2 $. 
It is evident that $G_b$ and $G_s$ correspond to the contributions with and without translation symmetry, respectively.
$G_s$ arises due to the Dirichlet boundary condition. 
On the other hand, in the presence of a boundary mass term, the boson propagator is
\begin{align}
	\label{eq:boson-prop}
     & D(k, x, x') = D_b(k, x, x') + w D_s(k, x, x') , \\
     & D_b(k, x, x') = \frac{e^{-q|x-x'|}}{2q} , 
     ~D_s(k, x, x') = \frac{e^{-q(x+x')} }{2q }.
\end{align}
The coefficient $w=\frac{q-c}{q+c}$ distinguishes the boundary conditions. 
In particular, the limit $c \rightarrow \infty$ corresponds to the Dirichlet boundary condition, while $c = 0$ corresponds to the Neumann boundary condition~\cite{diehl1981field,diehl1983multicritical,domb1986phase}. 
In these limits, the coefficient simplifies to $w=-1$ ($w=1$) in the Dirichlet (Neumann) boundary condition.

\paragraph{Renormalization group}\ssll
The propagators in~\eqref{eq:fermion-prop} and~\eqref{eq:boson-prop} serve as the building block for the RG analysis. 
In both $G_s$ and $D_s$, the dependence on the coordinate perpendicular to the boundary is exponentially suppressed when either $x$ or $x'$ is deep in the bulk, i.e., $x, x' \gg 1$. 
This suppression indicates that boundary effects do not contribute to the bulk RG equations. Consequently, the bulk RG equations remain identical to those of the chiral Ising universality class. 
For completeness, we present the one-loop RG equations in Supplemental Material at the fixed point, $\lambda^\ast =\frac{8\pi^2\left[(3-2N)+\sqrt{4N ^2+132N+9}\right]}{3(2N +3)}\epsilon$, $g^\ast =\frac{2\pi\sqrt{\epsilon}}{\sqrt{3/2+N}}$, corresponding to the well known chiral Ising universality class. 

The presence of a boundary introduces new critical exponents for the boundary fields. 
In particular, the two leading boundary operators are the Dirac fermion $\hat \psi \equiv \psi(x=0)$ and the scalar boson at the boundary. 
Here, we use a hat to denote boundary fields.
For the boundary scalar boson field, it is essential to distinguish between the Dirichlet and Neumann boundary conditions. Under the Dirichlet boundary condition, the field value is fixed to zero, $\hat \phi = 0$, making $\partial \hat \phi \equiv \partial_x \hat \phi$ the leading boundary scalar operator. 
In contrast, for the Neumann boundary condition, the leading boundary scalar field is $\hat \phi$ itself. 
Additionally, an extra relevant operator arises, corresponding to the boundary mass term $\hat \phi^2$. %

\begin{figure}
    \centering
    \subfigure[]{\includegraphics[width=0.22\linewidth]{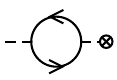}}~~
    \subfigure[]{\includegraphics[width=0.22\linewidth]{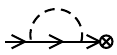}}~~ 
    \subfigure[]{\includegraphics[width=0.24\linewidth]{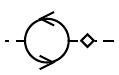}
    }~~
    \subfigure[]{\includegraphics[width=0.2\linewidth]{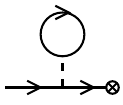}
    }
    \caption{Feynman diagrams relevant to the boundary field.
    The dashed (solid) line represents the boson (fermion) propagator. 
    The vertex $\otimes$ represents the boundary boson or fermion field and $\diamond$ represents the boundary mass term. 
    (a) Correction to the boundary boson. (b) Correction to the boundary fermion.
    (c) Correction to the boundary mass term. (d) Tadpole diagram for the correction to the boundary fermion.
    }
    \label{fig:feynman}
\end{figure}

We first examine the ordinary transition that corresponds to the Dirichlet boundary condition for the scalar field. 
The boundary critical exponent for a pure scalar theory has been previously determined, so our focus here is on the additional contribution arising from the presence of Dirac fermions. 
We introduce the boundary RG factors $Z_{\partial\hat \phi}$ and $Z_{\hat \psi}$ for the fermion field and boson fields, respectively, as follows, $\partial\hat \phi = \sqrt{Z_{\phi} Z_{\partial\hat \phi}} \partial \hat \phi_{\rm R }$ and $\psi = \sqrt{Z_{\psi} Z_{\hat \psi}}  \hat \psi_{\rm R}$.
Here $Z_\phi = 1-\frac{Ng^2}{4\pi^2\epsilon}$ and $Z_\psi = 1-\frac{g^2}{16\pi^2\epsilon}$ (at the one-loop order) are the RG factors of the bulk chiral Ising class. 

Consider the two-point function $\langle \phi \partial \hat \phi \rangle $. 
The Yukawa coupling generates a one-loop Feynman diagram for the boundary scalar field, illustrated in Fig.~\ref{fig:feynman}(a), which takes the form
\begin{equation}
	\label{eq:boson2}
    \int dx_1 dx_2 D(p,x,x_1)  G_2(p,x_1, x_2) [\partial_{x'} D(p, x_2, x')]_{x'=0}, \qquad 
\end{equation}
where $G_2$ represents the fermion bubble given by 
\begin{eqnarray}
	\label{eq:fermi-bubble}
     G_2(p, x_1, x_2) &=& \int d^{d-1}y e^{-i p  y} \\
     && \times \Tr\left[ \gamma^5 G(y, x_1, x_2) \gamma^5 G(-y, x_2, x_1) \right]. \nn
\end{eqnarray}
To interpret this expression, we start with a real-space representation, assigning the two vertices in Fig.~\ref{fig:feynman}(a) to the coordinates $(y, x_1)$ and $(0, x_2)$, respectively. 
We then treat the fermion bubble as an effective propagator rather than a closed loop and perform a Fourier transformation along the hyperspace direction to obtain the mixed representation. 
This precisely corresponds to the formulation in~\eqref{eq:fermi-bubble}. 
After incorporating this fermion bubble, integrating over $x_1$ and $x_2$  leads to a singularity, as detailed in Supplemental Material. 
The result can be summarized as:
\begin{multline}
    \langle \phi(p,x)\partial_x \phi(p',0)\rangle \\
    = (2\pi)^{d-1}\delta(p+p')e^{-px}\left(1+\frac{\lambda}{32\pi^2\epsilon}-\frac{Ng^2}{8\pi^2\epsilon} \right),
\end{multline}
where the Dirac $\delta$ function should be understood as componentwise. 
Inside the bracket, the first term corresponds to the tree-level contribution, the second term arises from boson self-interactions, and the third term represents the correction from the Yukawa coupling. 
From this, the RG factor for the boundary scalar field reads $ Z_{\partial\hat\phi}=1+\frac{\lambda}{16\pi^2\epsilon}+\frac{Ng^2}{4\pi^2\epsilon}$. 

Next, we analyze the two-point function $\langle \psi \hat {\bar \psi} \rangle$ to determine the RG factor for the boundary fermion field. 
The corresponding Feynman diagram, illustrated in Fig.~\ref{fig:feynman}(b), leads to the expression: $\int dx_1 dx_2 G(p,x,x_1) \tilde G_2(p,x_1, x_2)  G(p, x_2, 0)$ with
\begin{multline}
	\label{eq:fermi-boson}
     \tilde G_2(p, x_1, x_2) = \\
     -\int d^{d-1}y e^{-i p  y}  \gamma^5 G(y, x_1, x_2) \gamma^5 D(y, x_1, x_2).
\end{multline}
The reasoning behind this expression follows similarly from the analysis of the fermion bubble, and the RG procedure is also analogous. 
Therefore, we leave the detailed derivation to Supplemental Material and present the result for the RG factor, $Z_{\hat\psi} = 1-\frac{g^2}{16\pi^2\epsilon}$. 
With the RG factors, namely, the scaling dimension of the boundary fields at the ordinary transition can now be determined as $\Delta_{\partial \hat \phi} = \frac{d}2 + \eta_{\partial \hat \phi}$ where $\eta_{\partial \hat \phi} = \frac12 \frac{{\rm d} \log {Z_{\partial \hat \phi}} }{{\rm d} \log \mu} $ and similar for the Dirac fermions.
The results are summarized in Table~\ref{tab:scaling}.

Now, we turn to the discussion of the special transition, where the boson field satisfies the Neumann boundary condition. 
This scenario introduces several modifications, including changes in the leading boundary boson field and the boundary mass term: 
The leading boundary boson field is $\hat \phi$ instead of $\partial \hat \phi$; 
the inclusion of the boundary mass term is necessary because, at the special transition, the boundary mass term is a relevant perturbation.  
To account for these effects, we introduce the RG factors for the boson field and the boundary mass term:  $\hat \phi = \sqrt{Z_\phi Z_{\hat \phi}} \hat \phi_{\rm R} $ and $\hat \phi^2 = Z_{\hat \phi^2} \hat \phi^2_{\rm R}$. 

The calculation of the RG factors for the leading boundary boson and fermion fields follows the same approach as in the ordinary transition. The results are summarized as $ Z_{\hat\phi}=1+\frac{\lambda}{16\pi^2\epsilon}-\frac{Ng^2}{4\pi^2\epsilon}$ and $ Z_{\hat\psi} = 1-\frac{3g^2}{16\pi^2\epsilon}$.
Next, we discuss the evaluation of the RG factor $Z_{\hat \phi^2}$.
It can be obtained from the correlation function, $\langle \phi \phi \frac12 \hat \phi^2 \rangle$. 
The pure boson contribution is detailed in Supplemental Material.
Here, we focus on the one-loop diagram contributed from the Yukawa coupling, as shown in Fig.~\ref{fig:feynman}(c): $ \int dx_1 dx_2 D(p,x,0)D(p',0,x_1) G_2(p',x_1, x_2) D(p',x_2,x')$,
where $G_2(p,x_1, x_2)$ is the fermion bubble given in~\eqref{eq:fermi-bubble}. 
The divergence is evaluated in detail in Supplemental Material, and summarized as
\begin{align}
    &\langle \phi(p, x) \phi(p',x') \frac12 \phi^2(P,0) \rangle   = \\
    &(2\pi)^{d-1}\delta(P + p + p') \frac{e^{-p x-p' x'}}{p p'} \left(1-\frac{\lambda}{16\pi^2\epsilon}-\frac{3Ng^2}{8\pi^2\epsilon}\right),\nn
\end{align}
where $\phi^2(P,0) = \int d^{d-1} y e^{-i P y} \phi^2(y,0)$. 
From this, we obtain the RG factor for the boundary mass term: $Z_{\hat\phi^2}= 1-\frac{\lambda}{16\pi^2\epsilon}-\frac{3Ng^2}{8\pi^2\epsilon}$.
With the RG factors, the scaling dimension of the boundary fields at the special transition is determined similarly as before and summarized in Table~\ref{tab:scaling}.

\paragraph{Other variants of GNY universality class}\ssll
Our methodology can be extended to other variants of the GNY universality class. 
The chiral Heisenberg universality class describes a quantum phase transition from a Dirac semimetal to an antiferromagnetic insulator, where the order parameter is an $O(3)$ vector field~\cite{sorella2012absence,assaad2013pinning}:
\begin{align}
    S = \int_{\mathcal M} d^d x \Bigg( & \sum_j \bar\psi_j \gamma^\mu \partial_\mu \psi_j + \frac12 \sum_{i=1}^3(\partial \phi_i)^2  \nn \\
     & + \frac{\lambda}{4!} \Big( \sum_{i=1}^3 \phi_i^2 \Big)^2  -i g \sum_{i=1}^3 \sum_j  \phi_i \bar\psi_j  \gamma^5 s^i \psi_j \Bigg)\,,
\end{align}
where $\psi_j$ and $\phi_i$ denote the Dirac fermion and the vector boson fields, respectively.
The parameters $\lambda$ and $g$ represent the boson self-interaction and the Yukawa coupling, respectively.
The $\gamma$ matrices satisfy $\{\gamma^\mu, \gamma^\nu\} = 2 \delta^{\mu\nu}$ while $s^i$ are the Pauli matrices representing the spin degrees of freedom~\footnote{More precisely, the gamma matrices take the form $\gamma^\mu \otimes 1 $ and the spin matrices are $1 \otimes s^i$.}.
Notably, the inclusion of spin degrees of freedom results in an eight-component Dirac fermion. Furthermore, as before, we generalize the Dirac fermion to have $N$ flavors. 

The boundary condition for the Dirac fermion fields remains the same, $-\gamma^1 \psi|_\text{bdy} = \psi|_\text{bdy}$. 
This boundary condition can still be derived from a honeycomb lattice model with armchair boundaries. 
More importantly, because the $SU(2)$ spin rotation symmetry is independent of the presence of open boundaries, the boundary condition for the vector boson fields must also be compatible with spin rotation symmetry. 
As a result, the propagator for the vector boson fields takes the form $D_{ij}(k,x,x') = \delta_{ij} D(k,x,x')$, where $D(k,x,x')$ is given in~\eqref{eq:boson-prop}. 
With these propagators, the RG analysis follows the same procedure as in the chiral Ising universality class, and the final results are summarized in Table~\ref{tab:scaling}.

Finally, we discuss the role of the symmetry in the presence of open boundaries and the chiral XY universality class, which arises in the Kekul\'e valence-bond-solid transition in a 2D Dirac semimetal~\cite{li2017fermion}.
In previous cases, the presence of open boundaries does not affect the symmetry under consideration. 
In the case of the chiral Ising transition, the CDW breaks a reflection symmetry that interchanges two sublattices.
This reflection symmetry is intact in the presence of an armchair boundary. 
In the chiral Heisenberg transition, the spin rotation symmetry is also preserved in the presence of armchair boundaries. 
The fact that the symmetry remains intact ensures the vanishing of the Feynman diagram in Fig.~\ref{fig:feynman}(d).
To illustrate, consider a Yukawa coupling of the form: $\psi^\dag \Gamma \psi$, which transforms under a symmetry transformation $S$ as $S^\dag \Gamma S = - \Gamma$.  
This transformation property implies  $\Tr\left[\Gamma G\right] = 0 $, because $ \Tr\left[\Gamma G\right] = \Tr \left[\Gamma S G S^\dag \right] = -\Tr\left[\Gamma G\right] $. 
Here, we use the fact that the Green's function $G \sim \langle \psi \psi^\dag \rangle $~\footnote{This should be distinguished from the propagator in~\eqref{eq:fermion-prop}, which is presented in a relativistic form $G \sim \langle \psi \bar \psi \rangle$. However, the discussion does not rely on the specific representation.} is invariant under the symmetry transformation $S$. Notably, previous studies of the GNY model~\cite{giombi2021fermions,herzog2023fermions} did not account for this symmetry, leading to different results from  Fig.~\ref{fig:feynman}(d).

The Kekul\'e VBS transition provides a counterexample, because the presence of armchair boundaries explicitly breaks the translation symmetry under consideration. 
Consequently, the diagram in Fig.~\ref{fig:feynman}(d) does not vanish. 
Moreover, the loss of translation symmetry introduces a linear coupling to the VBS order parameter at the boundary. 
This suggests that the Kekul\'e VBS transition in the presence of an armchair boundary belongs to the normal universality class of the chiral XY universality class. 
On the contrary, the $U(1)$ symmetry at the transition to a pair-density wave (PDW) superconducting state remains intact~\cite{jian2015emergent}.
We leave the investigation of the PDW transition with possible boundary supersymmetric conformal field theory for a forthcoming work.

\paragraph{Concluding remarks}\ssll
To conclude, we have investigated the boundary criticality of the GNY model through explicit DQMC calculations and RG analysis. 
It would also be interesting to extend our DQMC simulation to larger system sizes and extract the boundary critical exponents numerically. 
We leave this simulation to the future work. 
More importantly, our systematic $4-\epsilon$ expansion can be extended to evaluate higher-order diagrams. 
For instance, the two-loop ``rainbow'' fermion self-energy diagram, as detailed in Supplemental Material, can be evaluated by iteratively performing our methods.
While our RG analysis primarily focuses on the ordinary and special transitions, an important direction for future work is to generalize the RG calculations to the extraordinary transition~\cite{bray1977critical,diehl1993critical,burkhardt1994ordinary,sun2025boundary}, where the bare fermion Green's function must incorporate a finite mass term induced by the scalar field in the chiral Ising transition. 
For the chiral Heisenberg transition, it would be particularly interesting to explore the possibility of an extraordinary-log transition~\cite{metlitski2020boundary,toldin2021boundary,hu2021extraordinary-log,padayasi2021the,zou2022surface,cuomo2023spontaneous,Zhang2022pott_afterreview,Sun2022villain_afterreview}, which would depend on the normal fixed point of the GNY model. 
Finally, the GNY universality class has been studied in various experimental settings~\cite{castro2009the,gomes2012designer,gutierrez2016imaging,bao2021experimental}. 
The boundary exponent for fermions can be probed using scanning tunneling microscopy at the system's edge. 

\vspace{1em}
\begin{acknowledgments}
{\it Acknowledgments---}%
The numerical mean field calculation was performed using the high-performance computational resources provided by the Louisiana Optical Network Infrastructure.
The DQMC simulations utilized ACES at Texas A\&M University High Performance Research Computing, through the allocation PHY250116 granted by the Advanced Cyberinfrastructure Coordination Ecosystem: Services and Support (ACCESS) program, which is supported by National Science Foundation grants \#2138259, \#2138286, \#2138307, \#2137603, and \#2138296.
This work is supported by a start-up fund (H.J., Y.G., and S.-K.J.) and a COR Research Fellowship (S.-K.J.) at Tulane University.

H.J. and Y.G. contributed equally to this work.
\end{acknowledgments}
	
\bibliography{references.bib}

\onecolumngrid
\setcounter{secnumdepth}{3}
\setcounter{equation}{0}
\setcounter{figure}{0}
\renewcommand{\theequation}{S\arabic{equation}}
\renewcommand{\thefigure}{S\arabic{figure}}
\renewcommand\figurename{Supplementary Figure}
\renewcommand\tablename{Supplementary Table}

\section*{Supplemental Material}

\section{Mean field theory of the honeycomb $t$-$V$ model}

\begin{figure}[t]
    \centering

    \subfigure[]{\includegraphics[width=0.28\linewidth]
    {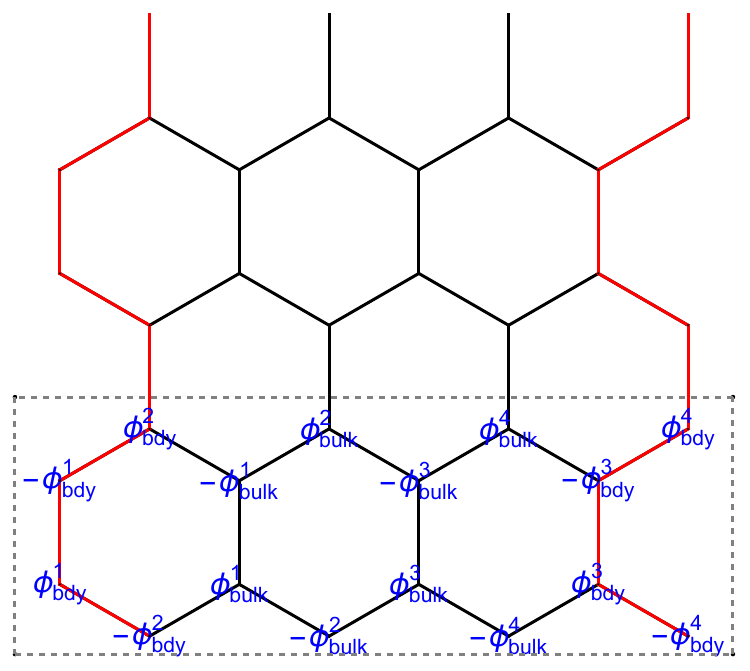}} \quad 
    \subfigure[]{\includegraphics[width=0.324\linewidth]
    {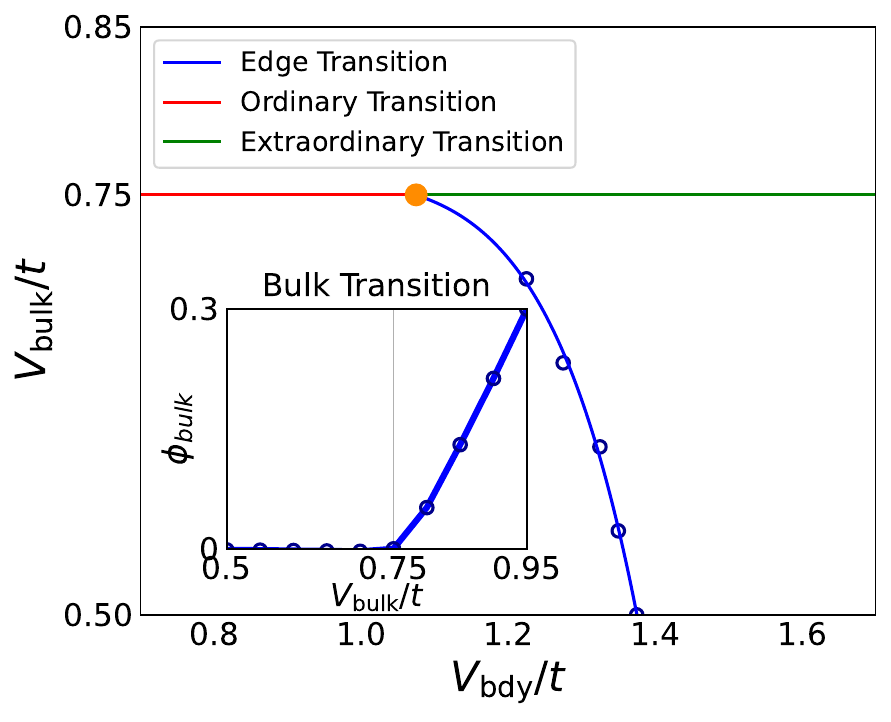}} \quad
    \subfigure[]{\includegraphics[width=0.33\linewidth]
    {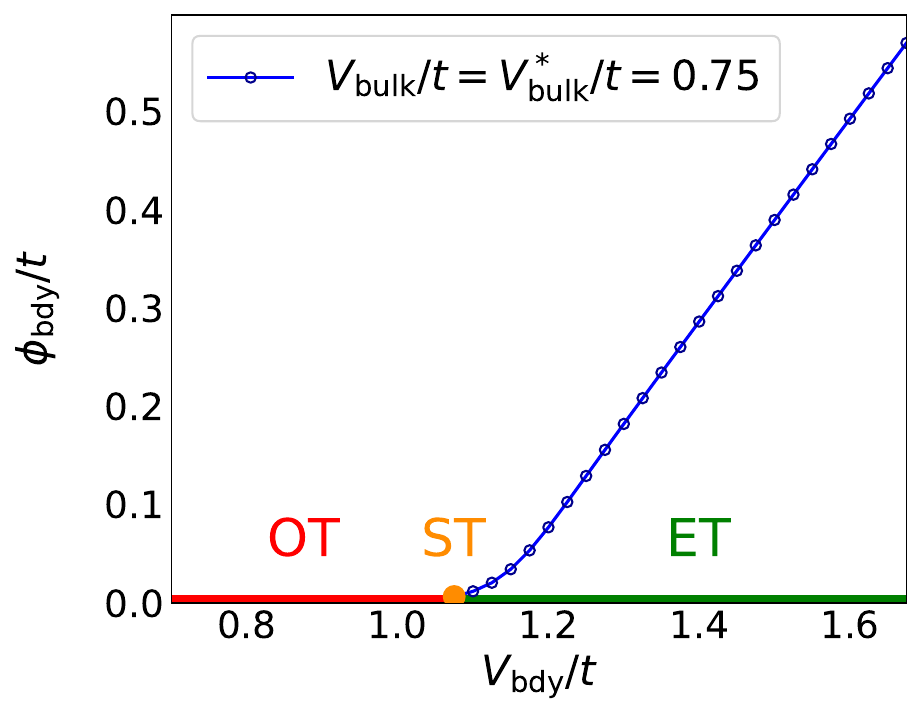}}

    \caption{(a) Illustration of the unit cell in a honeycomb lattice with armchair boundaries, along with the corresponding order parameters. 
    The interaction strength in black (red) bonds correspond to $V_\text{bulk}$ ($V_\text{bdy}$). 
    The gray dashed line delineates a unit cell. 
    $\phi$ denotes the CDW order parameter, where $\phi_\text{bdy}$ ($\phi_\text{bulk}$) represents the order on the boundary (in the bulk).
    The lattice is periodic along the $y$ axis, while the lattice length along the $x$ axis is chosen to be $L=4$ for an illustration. 
    In the mean field calculation, $L$ ranges from 25 to 49. (b) Phase diagram of lattice model \eqref{eq:lattice_sm}. 
    (c) Boundary order parameter as a function of the boundary interaction $V_\text{bdy}$ strength at the bulk critical point $V_\text{bulk}^*$. OT, ST, and ET represent ordinary, special, and extraordinary transition, respectively.  }
    \label{fig:mean_field_unit_cell}
\end{figure}

In this section, we detail the mean field theory calculation of the phase diagram.
We consider the Hamiltonian~\eqref{eq:lattice_sm} for spinless fermions on a honeycomb lattice with armchair boundaries:
\begin{align}    
    \label{eq:lattice_sm}
    H = - t \sum_{\langle \boldsymbol{ij} \rangle } ( c_{\boldsymbol i}^\dag c_{\boldsymbol j} + h.c.) + V_\text{bulk} \sideset{}{'}\sum_{\langle \boldsymbol{ij} \rangle } \left(n_{\boldsymbol i} - \frac12 \right) \left(n_{\boldsymbol j} - \frac12 \right)   + V_\text{bdy} \sideset{}{''}\sum_{\langle \boldsymbol{ij} \rangle } \left(n_{\boldsymbol i} - \frac12 \right) \left(n_{\boldsymbol j} - \frac12 \right),
\end{align}
where $V_\text{bulk}$ ($V_\text{bdy}$) represents the interaction strength on bulk (boundary) bonds, and $\langle \boldsymbol{ij}\rangle$ denotes the nearest-neighbor bonds. 
To compute the phase diagram, we use the following ansatz for the mean field Hamiltonian:
\begin{align} 
    \label{eq:mf_hamiltonian}
    H_\text{mf}\left[\{\phi_\text{bulk}^{\boldsymbol i},\phi_\text{bdy}^{\boldsymbol j}\}\right]=& -t\sum_{\langle \boldsymbol{ij}\rangle}(c_{\boldsymbol{i}}^\dagger c_{\boldsymbol{j}}+h.c.)+\sideset{}{'}\sum_{\boldsymbol i}\phi_\text{bulk}^{\boldsymbol i} (-1)^{s(\boldsymbol{i})} n_{\boldsymbol{i}}  +\sideset{}{''}\sum_{\boldsymbol i}\phi_\text{bdy}^{\boldsymbol i} (-1)^{s(\boldsymbol{i})} n_{\boldsymbol{i}}, 
\end{align}
Here, $s(\boldsymbol{i}) = 0, 1$ at two sublattices, and the sum $\sum_{\boldsymbol i}'$ and $\sum_{\boldsymbol i}''$ extend over the sites on bulk and boundary, respectively. 
We use $\phi^{\boldsymbol{i}}_\text{bulk/bdy}$ to denote a charge density wave (CDW) order ansatz, instead of $\Delta_{\boldsymbol{i}}$, in order to clearly distinguish our mean field results from those obtained in determinant quantum Monte Carlo (DQMC) simulations. 
Due to the symmetry of the armchair lattice, the order will still respect the translation symmetry along the $y$ axis while breaking the translation symmetry along the $x$ axis.
This is illustrated in Fig.~\ref{fig:mean_field_unit_cell}, where the dashed box denotes a unit cell of the CDW order in a lattice with $L = 4$. 
Here $L $ represents the length along the $x$ direction.

The CDW order is determined by the optimization of the mean field ground state energy, as detailed in the following. 
Since the ground state of the mean field Hamiltonian (the mean field ground state) is a Gaussian state, its energy can be obtained by calculating the correlation matrix, whose elements are defined as 
\begin{eqnarray}
    \Gamma_{\boldsymbol i,\boldsymbol j}\equiv \langle c_{\boldsymbol{i}}^\dagger c_{\boldsymbol{j}}\rangle_\text{mf} \equiv \langle \Psi_\text{mf} | c_{\boldsymbol{i}}^\dagger c_{\boldsymbol{j}} | \Psi_\text{mf} \rangle,
\end{eqnarray}
where $\Psi_\text{mf}$ is the ground state of the mean field Hamiltonian~\eqref{eq:mf_hamiltonian}.
We focus on the sector at half-filling.
In the diagonalized basis, the diagonal elements of the correlation matrix follow the Fermi-Dirac distribution and at zero temperature we have:
\begin{align}
    \Gamma^D=\left(\begin{matrix}
        0 & 0 & s & 0 & 0\\
        0 & 0 & s & 0 & 0\\
        \vdots & \vdots &\ddots & 0 & 0\\
        0 & 0 & s & 1 & 0\\
        0 & 0 & s & 0 & 1
    \end{matrix}\right),
\end{align}
where $\Gamma^D$ is the correlation matrix in the diagonal basis. 
We have arranged the eigen-energy in descending order, which implies that the upper (lower) half of the diagonal element $\Gamma^D$ is zero (one) in the half-filling sector.  
Denoting the unitary transformation  that diagonalizes the mean field Hamiltonian as $U$
(i.e., $U d= c$, with $ c$ the normal basis and $ d$ the diagonalized basis), the element of the correlation matrix is
\begin{align}
    \Gamma_{\boldsymbol i,\boldsymbol j}=\langle c_{\boldsymbol i}^\dagger c_{\boldsymbol j}\rangle_\text{mf}= \sum_{\boldsymbol{\alpha}, \boldsymbol{\beta}} \langle d_{\boldsymbol\alpha}^\dagger U_{\boldsymbol i\boldsymbol\alpha}^*U_{\boldsymbol j\boldsymbol\beta}d_{\boldsymbol\beta}\rangle_\text{mf}= \sum_{\boldsymbol{\alpha}, \boldsymbol{\beta}} U_{\boldsymbol i\boldsymbol\alpha}^*U_{\boldsymbol j\boldsymbol\beta} \langle d_{\boldsymbol\alpha}^\dagger d_{\boldsymbol\beta}\rangle_\text{mf}=(U^*\Gamma^DU^T)_{\boldsymbol i,\boldsymbol j},
\end{align}
where in the last step, we used the fact that correlation matrix in the diagonalized basis only has diagonal components. 
With this, the mean field ground state energy can be written as,
\begin{eqnarray}
    E\left[\{\phi_\text{bulk}^{\boldsymbol i},\phi_\text{bdy}^{\boldsymbol j}\}\right] = \left< H \right>_\text{mf} &=& -t\sum_{\langle \boldsymbol i\boldsymbol j\rangle}\left(\Gamma_{\boldsymbol i,\boldsymbol j} + \Gamma_{\boldsymbol j,\boldsymbol i}\right) + V_\text{bulk} \sideset{}{'}\sum_{\langle \boldsymbol i\boldsymbol j\rangle}\left[\left(\Gamma_{\boldsymbol i,\boldsymbol i}-\frac12\right)\left(\Gamma_{\boldsymbol j,\boldsymbol j}-\frac12\right)-\Gamma_{\boldsymbol i,\boldsymbol j}\Gamma_{\boldsymbol j,\boldsymbol i}\right] \nn \\
    && + V_\text{bdy} \sideset{}{''}\sum_{\langle \boldsymbol i\boldsymbol j\rangle}\left[\left(\Gamma_{\boldsymbol i,\boldsymbol i}-\frac12\right)\left(\Gamma_{\boldsymbol j,\boldsymbol j}-\frac12\right)-\Gamma_{\boldsymbol i,\boldsymbol j}\Gamma_{\boldsymbol j,\boldsymbol i}\right],
\end{eqnarray}
where correlation matrix $\Gamma_{\boldsymbol j,\boldsymbol i}$ is the function of $\phi_\text{bulk/bdy}^{\boldsymbol i}$. 
We vary the CDW order to get the minimal of the mean field ground state energy:
\begin{align}
    \frac{\delta E\left[\{\phi_\text{bulk}^{\boldsymbol i},\phi_\text{bdy}^{\boldsymbol j}\}\right]}{\delta \phi_\text{bulk}^{\boldsymbol{ i}}} \Bigg|_{\phi_\text{bulk/bdy}=\phi_\text{bulk/bdy}^{*}} =  \frac{\delta E\left[\{\phi_\text{bulk}^{\boldsymbol i},\phi_\text{bdy}^{\boldsymbol j}\}\right]}{\delta \phi_\text{bdy}^{\boldsymbol{ j}}} \Bigg|_{\phi_\text{bulk/bdy}=\phi_\text{bulk/bdy}^{*}} = 0, 
\end{align}
where $\{\phi_\text{bulk}^{\boldsymbol{i}*},\phi_\text{bdy}^{\boldsymbol{j}*}\}$ denotes the CDW order parameters corresponding to the minimal energy. 

In our calculation, we focus on a lattice geometry with the length perpendicular to the armchair boundary is given by $L=3 \mathbb Z+1$, so that the free fermion is gapless~\cite{brey2006eletronic}.
We used $L = 25, 37, 49$ and a similar number for the momentum points along $y$ direction. 
Notice that the previous work~\cite{talkachov2023microscopic} studied the mean field theory describing superconductivity in a honeycomb lattice with boundaries, where the spatial distribution of the order parameter exhibits nontrivial behavior in the presence of boundaries.

Using mean field calculation, we obtain the phase diagram shown in Fig.~\ref{fig:mean_field_unit_cell}(b).  
The boundary criticality characterizes the effect of the boundary at the quantum critical point $V_\text{bulk}=V_\text{bulk}^*$. 
As shown in Fig.~\ref{fig:mean_field_unit_cell}(c), when we fix the bulk interaction $V_\text{bulk}=V_\text{bulk}^*$, the boundary order parameter $\phi_\text{bdy}$ becomes nonzero once the boundary interaction strength
$V_\text{bdy}$ exceeds a critical value $V_\text{bdy}^*$. 
For $V_\text{bdy}>V_\text{bdy}^*$, the boundary order parameter acquires a finite value.
Hence, the mean field calculation identifies the ordinary transition $V_\text{bdy} < V_\text{bdy}^\ast$, the special transition $V_\text{bdy} = V_\text{bdy}^\ast$, and the extraordinary transition $V_\text{bdy} > V_\text{bdy}^\ast$. 
This is fully consistent with the determinant quantum Monte Carlo simulaitons.

\section{Details of the determinant quantum Monte Carlo simulation}

In this section, we provide the basic information of our determinant quantum Monte Carlo (DQMC) simulations. The simulation is carried out with the finite-temperature DQMC capabilities provided by the ALF package \cite{assaad2020ALF}. 
At interacting bond $(\boldsymbol i,\boldsymbol j)$, a Hubbard-Stratonovich transformation is used with an auxiliary field $\phi_{\boldsymbol i\boldsymbol j}$. 
The subsequent integral is approximated using the 4-point Gaussian-Hermite quadrature with $\eta_{\boldsymbol i\boldsymbol j}$ and $\gamma_{\boldsymbol i\boldsymbol j}$ at the roots labeled by $l=\pm1,\pm2$,
\begin{equation}
e^{-V_{\boldsymbol i\boldsymbol j}(n_{\boldsymbol i}-\frac12)(n_{\boldsymbol j}-\frac12)\mathrm{d}\tau+\frac14 V_{\boldsymbol i\boldsymbol j} \mathrm{d}\tau}=\frac{1}{\sqrt{\pi}}
\int\mathrm{d}\phi_{\boldsymbol i\boldsymbol j}e^{-\phi^2_{\boldsymbol i\boldsymbol j}+\phi_{\boldsymbol i\boldsymbol j}(c^\dagger_{\boldsymbol i}c_{\boldsymbol j}+c^\dagger_{\boldsymbol j}c_{\boldsymbol i})\sqrt{\frac{V_{\boldsymbol i\boldsymbol j}\mathrm{d}\tau}{2}}}\approx
\frac14\sum_{l=\pm1,\pm2} \gamma_{\boldsymbol i\boldsymbol j}(l) e^{\eta_{\boldsymbol i\boldsymbol j}(l)(c^\dagger_{\boldsymbol i}c_{\boldsymbol j}+c^\dagger_{\boldsymbol j}c_{\boldsymbol i})\sqrt{\frac{V_{\boldsymbol i\boldsymbol j}\mathrm{d}\tau}{2}}}\,
.
\end{equation}
The decomposition ensures that the Hamiltonian is sign-problem-free~\cite{li2014fermion}. Indeed, the quadratic Hamiltonian at an arbitrary auxiliary field configuration is simply the kinetic term with modified hopping amplitudes. We scale the inverse temperature with the system size by $\beta t=5L_x$ for $L_x=7,10,13$, so that the system is effectively at zero temperature for the computed observables. 
We set $t\mathrm{d}\tau=0.05$. 
For each physical parameter set at a system size, the computation utilizes 16 Markov chains, and between 220 to 2200 back-and-forth sweeps in each chain. 
Frequent stabilization is used to ensure numerical stability~\cite{assaad2020ALF}.

\begin{figure}
	\includegraphics[width=0.18\linewidth]{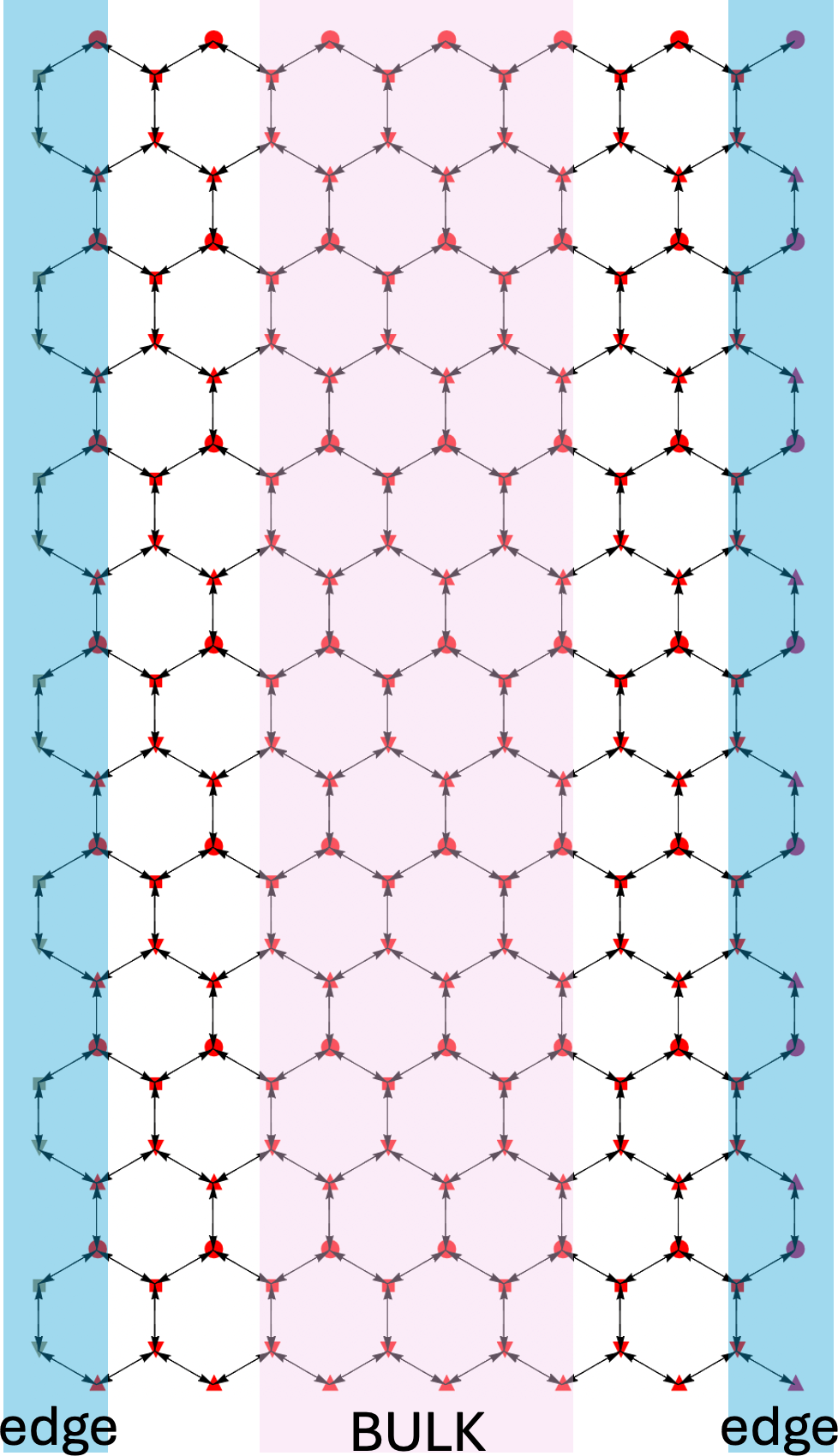}
	\caption{Illustration of the bulk and edge unit cells used in the $R_\text{bulk}$ and $R_\text{bdy}$ calculations in a lattice with $L_x=7$ and $L_y=14$, which is periodic in the $y$ direction. Only correlators between shaded sites are included in the corresponding calculation.}
	\label{fig:bulkedge-region}
\end{figure}

To compute the RG invariant quantities $R_\text{bulk}$ and $R_\text{bdy}$ in Eqs.~\eqref{eq:rginv} and \eqref{eq:ck} of the paper, we divide the lattice into bulk and edge regions separated by a single-unit-cell ribbon of lattice sites, as illustrated in Fig.~\ref{fig:bulkedge-region}. This reduces the finite-size effect, especially near the extraordinary transition due to the large $V_\text{bdy}$ felt by the inner ribbons of the lattice.

\addvspace{2em}

\section{Derivation of the boundary condition and propagators}

In this section, we derive the boundary condition for the Dirac fermion and the propagators.

\subsection{Boundary condition}
We give a brief derivation of the boundary condition for the Dirac fermion~\cite{brey2006eletronic}. 
At the ordinary and the special transition, the order parameter remains zero. 
Hence, the bare propagator will be given by the free model. 
$ H_0=-t\sum_{\langle\boldsymbol{ij}\rangle}(c_{\boldsymbol{i}}^\dagger c_{\boldsymbol{j}}+h.c.)$. 
It is well known that the free tight binding model, describing fermions hopping on a honeycomb lattice,
features two gapless points at ${K}$ and ${K}'$, respectively~\cite{castro2009the}. 
To be concrete, we choose the unit cell vectors to be $a_1 = ( 1, 0 )$ and $a_2 = (\frac12 , \frac{\sqrt 3}2) $, then consequently the two valleys are ${ K} = \frac{4\pi}3(1,0)$ and ${ K'} = \frac{4\pi}3(-1,0)$. 
Near these points, adopting $\mathbf{k}\mathbf{P}$ expansion, we arrive at a continuous model for the Dirac fermion, $\mathcal H=k_x\sigma^x\tau^z+k_y\sigma^y$ with the basis $\psi=(\psi_{A, K},\psi_{B,K},\psi_{A,K'},\psi_{B,K'})^{\rm T}$. 

Let us consider an armchair boundary along $y$ direction, as illustrated in Fig.~\ref{fig:mean_field_unit_cell}. 
With the translation symmetry breaking in $x$ direction, the eigenstate satisfies the Dirac equation,
\begin{align}
    (-i\partial_x\sigma^x\tau^z+k_y\sigma^y)\psi(x)= E \psi(x),
\end{align}
where $E$ denotes the eigen-energy. 
This equation is supplemented with proper boundary conditions at the armchair boundary.  
The boundary condition requires the wave function amplitudes to vanish at those sites (not shown) that would be connecting to the boundary (shown in the red color). 
We set the coordinates of these sites to be $x = 0$ for convenience, then we have $\Psi_{\sigma}(x=0) = 0 $ for $\sigma = A, B$, where $\Psi_{\sigma}(x)$ can be expressed in terms of the Dirac fermion,
\begin{eqnarray}
    \Psi_\sigma (x) = \psi_{\sigma,K}(x) e^{i K  x} + \psi_{\sigma,K'}(x) e^{i K'  x}. 
\end{eqnarray}
Therefore, in terms of the Dirac fermion, this boundary condition reads $\psi_{\sigma, K}(0) + \psi_{\sigma, K'}(0) = 0$, 
which can be expressed compactly in a matrix form,  
\begin{align}
    \tau^x \psi(x) = - \psi(x).
\end{align}
The boundary condition mixes two valleys because the fermion can scatter between valleys due to translation symmetry breaking at the boundary.

It will be convenient to work in a relativistic convention. 
To this end, we choose the following convention for the gamma matrix,
\begin{eqnarray}
\gamma^0=\tau^y\sigma^x, \gamma^1=\tau^x, \gamma^2=\tau^y\sigma^z, \gamma^3=\tau^z,  \gamma^5=\tau^y\sigma^y.
\end{eqnarray}
The free Dirac Lagrangian becomes $\mathcal L = \bar \psi \partial_\mu \gamma^\mu \psi$, with $\bar \psi = \psi^\dag \gamma^0$, and
the boundary condition becomes $\gamma^1\psi(0)=- \psi(0)$.

There is a more universal way to determine the boundary condition, without relying on the free model. 
Suppose the armchair boundary condition is given by a matrix $M$. 
This matrix should obey a few properties:
1) it anticommutes with the current in the $x$ direction because the perpendicular current vanishes at the boundary, 2) it commutes with the current in the $y$ direction, as the translation is preserved in the $y$ direction, 3) it commutes with the mirror symmetry $y \rightarrow -y$ because the armchair boundary respects this mirror symmetry, 4) it further commutes with the particle-hole symmetry.
Hence, we have 
\begin{eqnarray}
    \{ M, \tau^z \sigma^x \} = 0, \quad [M, \sigma^y ] = 0, \quad [M, \sigma^x] =0 , \quad [M, \sigma^x K] = 0,
\end{eqnarray}
where $K$ denotes complex conjugation. 
It is easy to check that the only matrix with these properties is $\gamma^1 = \tau^x$.
One can show that this boundary condition actually respects conformal symmetry~\cite{brillaux2023surface}. 

The above analysis gives the Dirichlet boundary condition for a semi-infinite geometry with an armchair boundary at $x=0$. 
It is enough for our RG calculation.
While, in our mean field calculation, the ribbon geometry has another armchair boundary.    
More specifically, if the honeycomb lattice has a finite length $L$ in $x$ direction, the boundary conditions require the Dirac field to satisfy
\begin{eqnarray}
    && \psi_{\sigma, K}(0)+ \psi_{\sigma, K'}(0)=0,\\
    && \psi_{\sigma, K}\left(L+ \frac12 \right)e^{i \frac{4\pi}3 \left( L + \frac12 \right)} + \psi_{\sigma, K'}\left(L + \frac12 \right) e^{- i \frac{4\pi}3 \left( L + \frac12 \right)} =0,
\end{eqnarray}
for the Dirichlet condition at the two armchair boundaries. 
It has been shown that with this boundary condition, the eigen-energy is~\cite{brey2006eletronic} 
\begin{eqnarray}
    E_n = \pm \sqrt{k_n^2 + k_y^2}, \quad k_n = 2\pi \left( \frac13 - \frac{n}{2L+1} \right), \quad n = 0,...,L-1. 
\end{eqnarray}
For $L=3 \mathbb Z +1$, the spectrum is gapless at $k_n = (2L+1)/3$.
Our mean field calculation has focused on these special lengths to better characterize the gapless states in the thermodynamic limit.

\subsection{Propagator of Dirac fermion}
With the Lagrangian and the boundary condition derived from the previous subsection, we calculate the bare propagator in this subsection. 
For generality, we consider a 3+1-dimensional Dirac fermion by including the $k_z$ direction.
The propagator satisfies the following different equation,
\begin{align}
\label{propagator}
(i\omega\gamma^0+\gamma^x\partial_x+ik_y\gamma^2+ik_z\gamma^3)G(k,x,x')=\delta(x-x'),
\end{align}
where $k$ denotes the momentum along the directions with translational symmetry. %
Expanding the matrix multiplication, one can show that \eqref{propagator} can be written as 
\begin{align}
    i\tau k_zG_{\tau\tau}^{\sigma\sigma}+(\partial_x+\tau\sigma k_y)G_{\bar\tau\tau}^{\sigma\sigma}+\tau\omega G_{\bar\tau\tau}^{\bar\sigma\sigma}&=\delta(x-x')\label{green_1}, \\
    i\tau k_zG_{\tau\tau}^{\bar\sigma\sigma}+(\partial_x-\tau\sigma k_y)G_{\bar\tau\tau}^{\bar\sigma\sigma}+\tau\omega G_{\bar\tau\tau}^{\sigma\sigma}&=0\label{green_2}, \\
    -i\tau k_zG_{\bar\tau\tau}^{\sigma\sigma}+(\partial_x-\tau\sigma k_y)G_{\tau\tau}^{\sigma\sigma}-\tau\omega G_{\tau\tau}^{\bar\sigma\sigma}&=0\label{green_3}, \\
    -i\tau k_zG_{\bar\tau\tau}^{\bar\sigma\sigma}+(\partial_x+\tau\sigma k_y)G_{\tau\tau}^{\bar\sigma\sigma}-\tau\omega G_{\tau\tau}^{\sigma\sigma}&=0\label{green_4}, 
\end{align}
where $\sigma = \pm$, $\bar \sigma = - \sigma$, $\tau = \pm$, $\bar \tau = - \tau$ denote the sublattice and valley space, respectively.
From \eqref{green_3} and \eqref{green_4}, we obtain
\begin{align}
    G_{\bar\tau\tau}^{\sigma\sigma}=\frac{(\partial_x-\tau\sigma k_y)G_{\tau\tau}^{\sigma\sigma}-\tau\omega G_{\tau\tau}^{\bar\sigma\sigma}}{i\tau k_z},\qquad G_{\bar\tau\tau}^{\bar\sigma\sigma}=\frac{(\partial_x+\tau\sigma k_y)G_{\tau\tau}^{\bar\sigma\sigma}-\tau\omega G_{\tau\tau}^{\sigma\sigma}}{i\tau k_z}.
\end{align}
Plugging them back into \eqref{green_1} and \eqref{green_2}, we arrive at
\begin{align}
    (\partial_x^2-q^2)G_{\tau\tau}^{\sigma\sigma}&=i\tau k_z\delta(x-x'),\\
    (\partial_x^2-q^2)G_{\tau\tau}^{\bar\sigma\sigma}&=0,
\end{align}
where we defined $q^2=\omega^2+k_y^2+k_z^2$. 
The solutions to these equations are then given by:
\begin{align}
    G_{\tau\tau}^{\sigma\sigma}(k, x,x')&=C_{\tau\tau}^{\sigma\sigma}(k,x')e^{-qx}+\frac{i\tau k_z}{2q}e^{-q|x-x'|}, \label{ansatz1} \\
    G_{\tau\tau}^{\bar\sigma\sigma}(k, x,x')&=D_{\tau\tau}^{\bar\sigma\sigma}(k, x')e^{-qx}, \label{ansatzs}
\end{align}
where $C_{\tau\tau}^{\sigma\sigma}(k, x')$ and $D_{\tau\tau}^{\bar\sigma\sigma}(k, x')$ are functions of $x'$ to be determined.
The boundary condition leads to $-\gamma^1 G(k,0,x)=G(k,0,x)$. 
Explicitly, this yield
\begin{align}
    -G_{\bar\tau\tau}^{\sigma\sigma}(k, 0,x)=G_{\tau\tau}^{\sigma\sigma}(k,0,x),\qquad -G_{\bar\tau\tau}^{\bar\sigma\sigma}(k,0,x)=G_{\tau\tau}^{\bar\sigma\sigma}(k,0,x) .\label{boundary_equation}
\end{align}
Hence, solving simultaneously the equations \eqref{green_1}--\eqref{green_4} and \eqref{ansatz1}--\eqref{boundary_equation}, we obtain the solutions
\begin{align}
    G_{\tau\tau}^{\sigma\sigma}(k,x,x')&=\frac{i\tau k_z}{2q}e^{-q|x-x'|}+\left(-\frac{1}{2}+\frac{\sigma\tau k_y}{2q}\right)e^{-q(x+x')},\\
    G_{\tau\tau}^{\bar\sigma\sigma}(k,x,x')&=\frac{\omega\tau}{2q}e^{-q(x+x')},\\
    G_{\bar\tau\tau}^{\sigma\sigma}(k,x,x')&=\left[-\frac{\tau\sigma k_y}{2q}-\frac{1}{2}{\rm sgn}\left(x-x'\right)\right]e^{-q|x-x'|}-\frac{i\tau k_z}{2q}e^{-q(x+x')},\\
    G_{\bar\tau\tau}^{\bar\sigma\sigma}(k,x,x')&=-\frac{\tau\omega}{2q}e^{-q|x-x'|}.
\end{align}

In matrix form, the propagator can be compactly expressed as
\begin{align} \label{eq:fermion_propagator}
    &G(k,x,x')=G_b(k,x,x')+G_s(k,x,x'), \\
    &G_b(k,x,x')=\left[\frac{i \slashed k}{2q}-\frac{\gamma^1}{2}{\rm sgn}\left(x-x'\right)\right]e^{-q|x-x'|}, \\
    &G_s(k,x,x')=-\gamma^1\left(\frac{i \slashed k}{2q}+\frac{\gamma^1}{2}\right)e^{-q(x+x')},
\end{align}
where $\slashed k=k_\mu \gamma^\mu$ for $\mu \ne 1$.
$G_b$ and $G_s$ correspond to contributions with and without translational symmetry, respectively.

To simplify the calculation of the fermion loop, we also transform the propagator to the real space via Fourier transformation. 
The Fourier transformation in $d-1$ flat space read
\begin{align}
    \int d^{d-1}y\frac{e^{iq y}}{(y^2+x^2)^\Delta}=\frac{2^{\frac{d+1}{2}-\Delta}\pi^{\frac{d-1}{2}}}{\Gamma(\Delta)}\left(\frac{q}{|x|}\right)^{\Delta-\frac{d-1}{2}}K_{\Delta-\frac{d-1}{2}}(q|x|).\label{fourier}
\end{align}
where $K$ is the modified Bessel function. This transformation yields
\begin{align}
    G_b(y,x,x')&=\frac{\Gamma(\frac{d}{2})}{2\pi^{d/2}}\frac{-\slashed y-(x-x')\gamma^1}{(y+(x-x')^2)^{d/2}},\\
    G_s(y,x,x')&=\frac{\Gamma(\frac{d}2)}{2\pi^{d/2}}\gamma^1\frac{- \slashed y +(x+x')\gamma^1}{(y^2+(x+x')^2)^{d/2}},
\end{align}
where $y$ denotes the coordinates in the direction with translational symmetry.

For completeness, we also present the boson propagator in real space,
\begin{align}
    D_b(y,x,x')&=\frac{\Gamma(\frac{d-2}{2})}{4\pi^{d/2}}\frac{1}{\left(y^2+(x-x')^2\right)^{\frac{d-2}{2}}}, \\
    D_s(y,x,x')&=\frac{\Gamma(\frac{d-2}{2})}{4\pi^{d/2}}\frac{1}{\left(y^2+(x+x')^2\right)^{\frac{d-2}{2}}}. 
\end{align}

\subsection{Method of image}

In addition to the detailed derivation of the propagator, one can use the method of image~\cite{nishioka2022method} to obtain the propagator in the Dirichlet and Neumann boundary conditions. 
According to the method of image, the propagator of a scalar in a semi-infinite space $\mathcal M$ is 
\begin{eqnarray}
    \langle \phi(x_1^\mu) \phi(x_2^\mu) \rangle_{\mathcal M} =    \langle \phi(x_1^\mu) \phi(x_2^\mu) \rangle_{\mathcal M'} + w \langle \phi(\bar x_1^\mu) \phi(x_2^\mu) \rangle_{ \mathcal M'},
\end{eqnarray}
where $\bar x^\mu$ is the image of $x^\mu$ by a reflection across the boundary plane $\partial \mathcal M$, and $w=1$ ($w=-1$) when $\phi$ satisfied Neumann (Dirichlet) boundary condition respectively. 
$\langle  \rangle_{\mathcal M}$ ($\langle  \rangle_{\mathcal M'}$) denotes the propagator in the semi-infinite (infinite) space. 
Hence, the propagator of the boson is 
\begin{eqnarray}
	\label{eq:boson_propagator_general}
    D(k, x, x') = D_b(k, x, x') + w D_b(k, \bar x, x'), \quad D_b(k, x, x') = \frac{e^{-q|x-x'|}}{2q} , 
\end{eqnarray}
where $q^2 = \sum_{\mu \ne 1} k_\mu^2$ and $D_b $ is the propagator in the infinite space. 
$\bar x= -x$ is the image point with respect to the boundary, and $k$ denotes the momentum in the space parallel to the boundary.
For spinor field, we can generalize the method of image 
\begin{eqnarray}
     \langle \psi(x_1^\mu) \bar \psi(x_2^\mu) \rangle_{\mathcal M} =    \langle \psi(x_1^\mu) \bar \psi(x_2^\mu) \rangle_{\mathcal M'} + w \gamma^1 \langle \psi(\bar x_1^\mu) \psi( x_2^\mu) \rangle_{ \mathcal M'}
\end{eqnarray}
where $\gamma^1$ is the reflection matrix across the $x$ direction. 
Hence, the propagator of the Dirac fermion in the semi-infinite space is 
\begin{eqnarray}
	\label{eq:fermion_propagator_general}
    G(k, x, x') &=&  G_b(k, x, x') + w \gamma^1 G_b(k, \bar x, x'), \quad  
    G_b(k,x,x') = \left(\frac{i\slashed{k}}{2q}-\frac{\gamma^1}{2} \sgn(x-x') \right)e^{-q|x-x'|}. 
\end{eqnarray}
For the Dirichlet boundary condition $w = -1$, it is consistent with the result from the last subsection~\eqref{eq:fermion_propagator}. 

\section{Renormalization group calculation for the chiral Ising universality class}

In this section, we detail the RG calculation for the chiral Ising universality class. 
The action for the chiral Ising universality class reads, 
\begin{eqnarray}
    S &=& \int_\mathcal{M}d^dx \Bigg( \sum_j\bar\psi_j\gamma^\mu\partial_\mu\psi_j 
    + \frac12 (\partial_\mu \phi)^2 + \frac{\lambda}{4!} \phi^4  -ig\sum_j\phi\bar\psi_j\gamma^5\psi_j \Bigg),
\end{eqnarray}
where $\psi_j$ represents the Dirac fermion fields and $\phi$ is the order parameter field. $\partial_\mu$ denotes derivatives regarding (imaginary) time $x_0 \equiv \tau$ and space $x_i$, $i=1,...,d-1$. 
$\gamma_\mu$ are gamma matrices: $\gamma^0=\tau^y\sigma^x$, $\gamma^1=\tau^x$, $\gamma^2=\tau^y\sigma^z$, $\gamma^3=\tau^z$, and $\gamma^5=\tau^y\sigma^y$, and $\bar \psi = \psi^\dag \gamma^0$. 
Note that, to make our theory more general, we extend the fermion sector to include $N$ flavors, denoted as $\psi_j$ for $j=1,...,N$. 
The parameters $\lambda$ and $g$ represent the quartic boson self-interaction and the fermion-boson Yukawa coupling, respectively. 
The field theory is located in a $d$-dimensional semi-infinite spacetime, $\mathcal M = \{ x_\mu | x_1>0 \}$ where the boundary is $\partial \mathcal M = \{ x_\mu| x_1=0 \}$. 
We denote the hyperspace parallel to the boundary as $y$ and the coordinate perpendicular to the boundary as $x$. 
The boundary condition for the Dirac fermion arises from the vanishing of the lattice fermion outside the lattice sites, enforcing a Dirichlet boundary condition. 
The boundary condition for the Dirac fermion field is $-\gamma^1 \psi|_{x=0} = \psi|_{x=0}$. 
The bulk $4-\epsilon$ RG analysis for the chiral Ising universality is well known~\cite{zerf2017four-loop}, here we present the RG factors in the one-loop order,
\begin{eqnarray}
    && Z_\phi=1-\frac{Ng^2}{4\pi^2\epsilon}, \quad 
    Z_\psi=1-\frac{g^2}{16\pi^2\epsilon}, \\
    && Z_g=1+\frac{(3+2N)g^2}{16\pi^2\epsilon}, \quad 
    Z_\lambda=1+\frac{Ng^2}{2\pi^2\epsilon}+\frac{3\lambda}{16\pi^2\epsilon}-\frac{N3g^4}{\lambda\pi^2\epsilon},
\end{eqnarray}
which are defined as
\begin{eqnarray}
    \phi = \sqrt{ Z_\phi} \phi_{\rm R}, \quad \psi = \sqrt{ Z_\psi} \psi_{\rm R}, \quad 
    g = \mu^{\epsilon/2} Z_g g_{\rm R}, \quad \lambda = \mu^{\epsilon} Z_\lambda \lambda_{\rm R}
\end{eqnarray}
where $\hat \phi_R$, $\hat \psi_R$, $g_{\rm R}$ and $\lambda_{\rm R}$ denote the renormalized quantities.
Consequently, the RG equation at the one-loop order reads,
\begin{eqnarray}
	\label{eq:chiral_ising}
    \frac{{\rm d}\lambda}{{\rm d} \log \mu} &=& - \epsilon\lambda +\frac{N\lambda g^2}{2\pi^2} + \frac{3\lambda^2}{16\pi^2} - \frac{3Ng^4}{\pi^2}, \\ 
    \frac{{\rm d}g}{{\rm d} \log \mu} &=& - \frac{\epsilon}{2}g + \frac{(3+2N)g^3}{16\pi^2}.
\end{eqnarray}
This leads to the chiral Ising fixed point at 
\begin{eqnarray}
    \lambda^\ast =\frac{8\pi^2\left[(3-2N)+\sqrt{4N ^2+132N+9}\right]}{3(2N +3)}\epsilon, \quad g^\ast =\frac{2\pi\sqrt{\epsilon}}{\sqrt{3/2+N}} .
\end{eqnarray}
Furthermore, the scaling dimensions for the boson and fermion are, respectively, given by
\begin{eqnarray}
    \Delta_{\phi} = 1-\frac{3}{6+4N}\epsilon , \quad 
    \Delta_{\psi} = \frac{3}{2}-\frac{5+4N}{12+8N}\epsilon. 
\end{eqnarray}
where $N$ is the flavor of the four-component Dirac fermion. 
In the following, we will calculate the boundary critical exponents at the ordinary and special transition.
The fermion satisfies the Dirichlet boundary condition at both transitions, so the bare fermion propagator is~\eqref{eq:fermion_propagator_general} with $w=-1$.

\subsection{Ordinary transition} 

At the ordinary transition, the boson field obeys the Dirichlet boundary conditions. 
Hence, the boson propagator is~\eqref{eq:boson_propagator_general} with $w=-1$.
The RG factors for the boundary fields are defined by~\cite{diehl1983multicritical},
\begin{eqnarray}
    \hat \phi = \sqrt{Z_\phi Z_{\partial \hat \phi}} \partial \hat \phi_{\rm R},   \quad  \hat \psi = \sqrt{Z_\psi Z_{\hat \psi}} \hat \psi_{\rm R},
\end{eqnarray}
where $\hat \phi_R$ and $\partial\hat \phi_R$ denote the renormalized fields.

As stated in the main text, to get the scaling dimension for $\partial \hat \phi$, we consider the correction to the correlation function $\langle \phi \partial \hat \phi \rangle$.
At the tree level, the correlation is given by
\begin{align}
    \langle\phi(p,x)\partial_x\hat\phi(p',0)\rangle= (2\pi)^{d-1} \delta(p+p') e^{-px}.
\end{align}
At the one-loop level, the correction from the fermions, illustrated in the Feynman diagram Fig.~2(a), is 
\begin{align}
	\label{eq:correction_to_boson}
    \int dx_1dx_2D(p,x,x_1) G_2(p,x_1,x_2)\left[\partial_xD(p,x_2,x) \right]_{x=0}.
\end{align}
where $G_2$ denotes the fermion loop,
\begin{align}
    G_2(p,x_1,x_2)=&-\int d^{d-1}y e^{-ip  y} {\rm Tr}\left[(-i\gamma^5)G(y,x_1,x_2)(-i\gamma^5)G(-y,x_2,x_1)\right]    \nonumber\\
    =&\frac{\Gamma(\frac{d}{2})^2}{\pi^d}\int d^{d-1}y e^{-ip  y} \left[\frac{1}{(y^2+(x_1-x_2)^2)^{d-1}}+\frac{1}{(y^2+(x_1+x_2)^2)^{d-1}}\right]\\
    \to&\frac{1}{8\pi^2}(2|x_1-x_2|^{-3+\epsilon}-q^2|x_1-x_2|^{-1+\epsilon})+\frac{1}{8\pi^2}(2(x_1+x_2)^{-3+\epsilon}-q^2(x_1+x_2)^{-1+\epsilon}),
\end{align}
where in the last line, we perform the Fourier transformation using~\eqref{fourier}, and then isolate the pole.  
To evaluate the divergences, we use the properties of the generalized function~\cite{diehl1983multicritical,gelfand1964generalized}. 
For any function $\varphi$, a generalized function $x^{\lambda}$ with ${\rm Re} \lambda > - n -1$ has the poles at $\lambda = -1, -2, ...$ given by~\cite{gelfand1964generalized}
\begin{eqnarray}
    \int_0^\infty dx\, x^\lambda \varphi(x) = \sum_{k=1}^n \frac{\varphi^{(k-1)}(0)}{(k-1)!(\lambda + k)}. 
\end{eqnarray}
In particular, the residue is $\varphi^{(k-1)}(0)/(k-1)!$ at $\lambda = -k$. 
Generalizing this result, we have
\begin{eqnarray}
    && \int dx_1 d x_2 \,(x_1+x_2)^{-3+\epsilon} \varphi(x_1, x_2) 
    = \frac1{2\epsilon}[\varphi^{(1,0)}(0,0) + \varphi^{(0,1)}(0,0)], \\
    && \int dx_1dx_2\,|x_1-x_2|^{-3+\epsilon} \varphi(x_1,x_2)= \frac1{\epsilon}\int_0^\infty dx[\varphi^{(2,0)}(x,x)+ \varphi^{(0,2)}(x,x)], \\
    && \int dx_1dx_2\,|x_1-x_2|^{-1+\epsilon} \varphi(x_1,x_2)= \frac2\epsilon \int _0^\infty dx \, \varphi(x,x),
\end{eqnarray}
where we used the shorthand notation
\begin{eqnarray}
    \varphi^{(n,m)}(x_1, x_2) = \frac{\partial^n}{\partial x_1^n} \frac{\partial^m}{\partial x_2^m} \varphi(x_1, x_2). 
\end{eqnarray}
And there are no poles in $\epsilon$ at $m=1$ for $(x_1+x_2)^{-1+\epsilon}$.

Using these results, we can evaluate the integration in~\eqref{eq:correction_to_boson} and obtain the divergence.
\begin{align}
    \int dx_1dx_2\,D(p,x,x_1) G_2(p,x_1,x_2)\left[\partial_xD(p,x_2,x) \right]_{x=0} = -\frac{e^{-px}}{8\pi^2\epsilon}.
\end{align}
Combining the result from the pure boson contribution~\cite{diehl1983multicritical}, we arrive at the correction to the boson correlation function,  
\begin{equation} \label{eq:Z_dphi}
    \langle\phi(p,x)\partial_x\phi(p',0)\rangle=(2\pi)^{d-1}\delta(p+p')e^{-px}\left(1+\frac{\lambda}{32\pi^2\epsilon}-\frac{Ng^2}{8\pi^2\epsilon} \right).
\end{equation}

Next, to obtain the scaling dimension of the boundary fermion $\hat \psi$, let's consider the correction to the correlation function $\langle \psi \hat \psi\rangle$.
The one-loop correction from the Feynman diagram Fig.~2(b) leads to
\begin{align}
    \int dx_1dx_2\,G(k,x,x_1)\tilde G_2(k,x_1,x)G(k,x_2,0),
\end{align}
where $\tilde G_2(k,x_1,x_2)$ is given by Fourier transformation
\begin{align} \label{eq:integration_fermion_boson}
    \tilde G_2(k,x_1,x_2)=\int d^{d-1}y \, e^{-i k y}(-i\gamma^5)G(y,x_1,x_2)(-i\gamma^5)D(y,x_1,x_2).
\end{align}
For computational convenience, we divide $\tilde G_2$ into the following four terms,
\begin{align}
    G_bD_b&=\frac{\Gamma(\frac{d}{2})\Gamma(\frac{d-2}{2})}{8\pi^d}\frac{-\slashed y -(x_1-x_2)\gamma^1}{(y^2+(x_1-x_2)^2)^{d-1}}, \\
    G_sD_s&=\frac{\Gamma(\frac{d}{2})\Gamma(\frac{d-2}{2})}{8\pi^d}\gamma^1\frac{-\slashed y +(x_1+x_2)\gamma^1}{(y^2+(x_1+x_2)^2)^{d-1}}, \\
    G_bD_s&=\frac{\Gamma(\frac{d}{2})\Gamma(\frac{d-2}{2})}{8\pi^d}\frac{-\slashed y-(x_1-x_2)\gamma^1}{(y^2+(x_1-x_2)^2)^{d/2}(y^2+(x_1+x_2)^2)^{d/2-1}}     ,\nonumber \\
    &=\frac{\Gamma(d-1)}{8\pi^d}\int_0^1 du \, u^{d/2-1}(1-u)^{d/2-2}\frac{-\slashed y-(x_1-x_2)\gamma^1}{(y^2+x_1^2+x_2^2+2(1-2u)x_1x_2)^{d-1}}\label{gbds} ,\\
    G_sD_b&=\frac{\Gamma(\frac{d}{2})\Gamma(\frac{d-2}{2})}{8\pi^d}\gamma^1\frac{-\slashed y+(x_1+x_2)\gamma^1}{(y^2+(x_1+x_2)^2)^{d/2}(y^2+(x_1+x_2)^2)^{d/2-1}}   ,\nonumber \\
    &=\frac{\Gamma(d-1)}{8\pi^d}\int_0^1 du \,u^{d/2-1}(1-u)^{d/2-2}\gamma^1\frac{-\slashed y+(x_1+x_2)\gamma^1}{(y^2+x_1^2+x_2^2-2(1-2u)x_1x_2)^{d-1}\label{gsdb}}    .
\end{align}
where in the last line of \eqref{gbds} and \eqref{gsdb}, we applied the Feynman parameterization and
$\slashed y \equiv y_\mu \gamma^\mu$ with $\mu \ne 1$. 
The Fourier transformation can be performed with the help of~\eqref{fourier}, 
\begin{align}
    \frac{-\slashed y-x\gamma^1}{(y^2+\tilde x^2)^{d-1}}\to\frac{\pi^2}{4}\tilde x^{-1+\epsilon} i\slashed{k} -\frac{\pi^2}{4}\left(\tilde x^{-3+\epsilon}-\frac{q^2}{2}\tilde x^{-1+\epsilon}\right)x\gamma^1, 
\end{align}
where $\slashed k=k_\mu \gamma^\mu$ for $\mu \ne 1$. 
We only list the terms that contribute to $\frac 1\epsilon$ divergence, relevant to the one-loop RG calculations. 

Now, we can use the result for the generalized function and then perform the integration in~\eqref{eq:integration_fermion_boson}. 
The contribution from $G_bD_b$ is
\begin{align}
    &\frac{\Gamma\left(\frac{d}{2}\right)\Gamma\left(\frac{d-2}{2}\right)}{8\pi^d} \int dx_1dx_2G(k,x,x_1)(-i\gamma^5)\left(\frac{\pi^2}{4}|x_1-x_2|^{-1+\epsilon}i\slashed k-\frac{\pi^2}{4}|x_1-x_2|^{-2+\epsilon}{\rm sgn}\left(x_1-x_2\right)\gamma^1\right) (-i\gamma^5) \nn \\
    & \times G(k,x_2,0)=-\frac{1}{16\pi^2\epsilon}\left[\frac{i\slashed k}{2q}-\frac{\gamma^1}{2}-\gamma^1\left(\frac{i\slashed k}{2q}+\frac{\gamma^1}{2}\right)\right]e^{-qx},
\end{align}
the contribution from $G_sD_s$ is
\begin{align}
    &\frac{\Gamma\left(\frac{d}{2}\right)\Gamma\left(\frac{d-2}{2}\right)}{8\pi^d}\int dx_1dx_2G(k,x,x_1)(-i\gamma^5)\left(\gamma^1\frac{\pi^2}{4}\left(x_1+x_2\right)^{-2+\epsilon}\gamma^1\right)(-i\gamma^5)G(k,x_2,0), \nonumber \\
    &=\frac{1}{32\pi^2\epsilon}\left[\frac{i\slashed k}{2q}-\frac{\gamma^1}{2}-\gamma^1\left(\frac{i\slashed k}{2q}+\frac{\gamma^1}{2}\right)\right]e^{-qx},
\end{align}
the contribution from $G_bD_s$ vanishes and
the contribution from $G_sD_b$ is
\begin{align}
    &\frac{\Gamma(d-1)}{8\pi^d}\int_0^1duu\int dx_1dx_2G(k,x,x_1)(-i\gamma^5)\left[\frac{\pi^2}{4}\left(x_1^2+x_2^2-2(1-2u)x_1x_2\right)^{-3/2+\epsilon/2}\gamma^1(x_1+x_2)\gamma^1\right] (-i\gamma^5) \nonumber \\
    &\times G(k,x_2,0)=\frac{1}{16\pi^2\epsilon}\left[\frac{i\slashed k}{2q}-\frac{\gamma^1}{2}-\gamma^1\left(\frac{i\slashed k}{2q}+\frac{\gamma^1}{2}\right)\right]e^{-qx}.
\end{align}

Hence, the total correction to the correlation function $\langle \psi \hat {\bar \psi} \rangle$ reads
\begin{align} \label{eq:Z_dpsi}
    \langle\psi(k,x)\bar\psi(k',0)\rangle=(2\pi)^{d-1}\delta(k-k')\left(1-\frac{3 g^2}{32\pi^2\epsilon}\right)\left[\frac{i\slashed k}{2q}-\frac{\gamma^1}{2}-\gamma^1\left(\frac{i\slashed k}{2q}+\frac{\gamma^1}{2}\right)\right]e^{-qx}.
\end{align}

With the corrections in~\eqref{eq:Z_dphi} and~\eqref{eq:Z_dpsi}, the RG factors are 
\begin{eqnarray}
    Z_{\partial\hat\phi}=1+\frac{\lambda}{16\pi^2\epsilon}+\frac{Ng^2}{4\pi^2\epsilon}, \qquad Z_{\hat\psi} = 1-\frac{g^2}{16\pi^2\epsilon}. 
\end{eqnarray}
Accordingly, the corresponding anomalous dimensions are
\begin{eqnarray}
    \eta_{\partial \hat \phi} = \frac12 \frac{{\rm d} \log {Z_{\partial \hat \phi}} }{{\rm d} \log \mu} = -\frac{\lambda}{32\pi^2}-\frac{Ng^2}{8\pi^2}, \quad \eta_{ \hat \psi} = \frac12 \frac{{\rm d} \log {Z_{ \psi}} }{{\rm d} \log \mu} = \frac{g^2}{32\pi^2}.
\end{eqnarray}
At the chiral Ising fixed point, we arrive at the results summarized in Table~\ref{tab:scaling}. 

\subsection{Special transition}
At the ordinary transition, the boson field obeys the Dirichlet boundary conditions. Hence, the boson propagator is~\eqref{eq:boson_propagator_general}
with $w = 1$. 
The RG factors for the boundary fields are defined by~\cite{diehl1983multicritical},
\begin{eqnarray}
    \hat \phi = \sqrt{Z_\phi Z_{\hat \phi}} \hat \phi_{\rm R}, \quad  \hat \phi^2 = Z_{\hat \phi^2} \hat \phi^2_{\rm R}, \quad  \hat \psi = \sqrt{Z_\psi Z_{\hat \psi}} \hat \psi_{\rm R},
\end{eqnarray}
where $\hat \phi_R$, $\hat \phi_R^2$, and $\hat \psi_R$ denote the renormalized fields.

To obtain the scaling dimension for $\hat \phi$, we consider the correction to the correlation
function $\langle \phi \hat \phi\rangle$.
The one-loop Feynman diagram in Fig.~2(a) gives,
\begin{align}
	\label{eq:correction_fermion}
    \int dx_1dx_2 D(p,x,x_1) G_2(p,x_1,x_2)D(p,x_2,0).
\end{align}
Following a similar procedure employed in \eqref{eq:correction_to_boson}--\eqref{eq:Z_dphi}, and combining the result from the pure boson contribution~\cite{diehl1983multicritical}, the divergence is 
\begin{eqnarray}\label{eq:Z_phi}
    \langle \phi(p, z) \phi(p', 0) \rangle = (2\pi)^{d-1} \delta(p+p') \frac{e^{-pz}}{p} \left( 1 + \frac{\lambda}{32\pi^2 \epsilon} -\frac{3Ng^2}{8\pi^2\epsilon} \right).
\end{eqnarray}

Next, as discussed in the main text, the special transition has another relevant field, $\frac{1}{2}\hat\phi^2$. 
To this end, we consider the correction to the correlation function $\langle \phi \phi \frac12 \hat \phi^2\rangle$.
For the Feynman diagram shown in Fig.~2(c), the contribution is identical to the fermion bubble correction to the boundary boson $\hat\phi$. 
Combining the correction from pure boson contribution~\cite{diehl1983multicritical}, we arrive at
\begin{align} \label{eq:Z_phi2}
     \langle\phi(p_1,x_1)\phi(p_2,x_2)\frac{1}{2}\hat{\phi}^2(P,0)\rangle=(2\pi)^{d-1}\delta(P+p_1+p_2)\frac{e^{-p_1x_1-p_2x_2}}{p_1p_2}\left(1-\frac{\lambda}{16\pi^2\epsilon}-\frac{3Ng^2}{8\pi^2\epsilon}\right).
\end{align}

Finally, for the boundary fermion $\hat \psi$, the correction to the correlation function $\langle \psi \bar \psi \rangle$ is given, according to the one-loop Feynman diagram Fig.~2(b), by
\begin{align}
    \int dx_1dx_2G(k,x,x_1)\tilde G_2(k,x_1,x_2)G(k,x_2,0),
\end{align}
where $\tilde G_2(k,x_1,x_2) = \int d^{d-1}y e^{-i k y}(-i\gamma^5)G(y,x_1,x_2)(-i\gamma^5)D(y,x_1,x_2) $.
The only difference is that the boson propagator satisfies the Neumann boundary condition.
The calculation follows the previous discussion at the ordinary transition, and the final expression for the correction to the correlation function $\langle \psi \hat \psi\rangle$ reads,
\begin{align} \label{eq:Z_psi}
    \langle\psi(k,x)\bar\psi(k',0)\rangle=(2\pi)^{d-1}\delta(k-k')\left(1-\frac{5g^2}{32\pi^2\epsilon}\right)\left[\frac{i\slashed k}{2q}-\frac{\gamma^1}{2}-\gamma^1\left(\frac{i\slashed k}{2q}+\frac{\gamma^1}{2}\right)\right]e^{-qx}.
\end{align}

With the corrections in~\eqref{eq:Z_phi},~\eqref{eq:Z_phi2} and~\eqref{eq:Z_psi}, the RG factors are 
\begin{eqnarray}
    Z_{\hat\phi}=1+\frac{\lambda}{16\pi^2\epsilon}-\frac{Ng^2}{4\pi^2\epsilon}, \quad Z_{\hat\phi^2}=1-\frac{\lambda}{16\pi^2\epsilon}-\frac{3Ng^2}{8\pi^2\epsilon}, \quad
    Z_{\hat\psi}= 1-\frac{3g^2}{16\pi^2\epsilon}. 
\end{eqnarray}
Accordingly, the corresponding anomalous dimensions are
\begin{eqnarray}
    \eta_{\hat \phi} = \frac12 \frac{{\rm d} \log {Z_{\hat \phi}} }{{\rm d} \log \mu} = -\frac{\lambda}{32\pi^2} + \frac{Ng^2}{8\pi^2}, \quad 
     \eta_{\hat \phi^2} =  \frac{{\rm d} \log {Z_{\hat \phi^2}} }{{\rm d} \log \mu} = \frac{\lambda}{16\pi^2}+ \frac{Ng^2}{8\pi^2},
     \quad \eta_{ \hat \psi} = \frac12 \frac{{\rm d} \log {Z_{ \psi}} }{{\rm d} \log \mu} = \frac{3g^2}{32\pi^2}.
\end{eqnarray}
At the chiral Ising fixed point, we arrive at the results summarized in Table~\ref{tab:scaling}.

\subsection{Extend to higher orders}
In this section, we briefly discuss how our method can be extended to higher orders.
\begin{figure}[t]
    \includegraphics[width=0.4\linewidth]{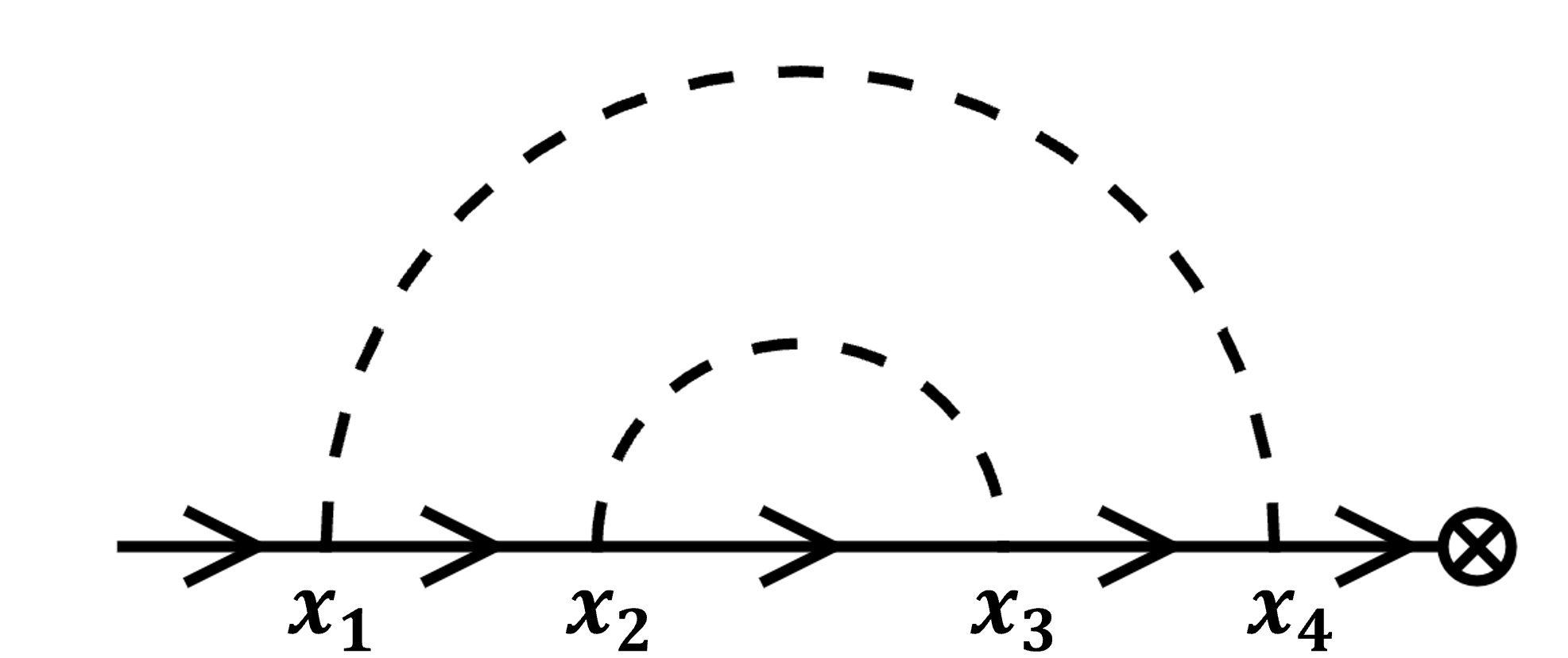}
    \caption{The two-loop ``rainbow'' diagram. The dashed (solid) line represents the boson (fermion) propagator. The vertex $\bigotimes$ denotes the boundary fermion.}
    \label{fig:rainbow_sm}
\end{figure}
In the following, we illustrate how our method can be directly generalized to higher orders by discussing the calculation of a two-loop Feynman diagram. 
Specifically, we consider the ``rainbow'' diagram, shown in Fig.~\ref{fig:rainbow_sm}, which contributes to the two-loop fermion self-energy for the boundary criticality within the chiral Ising universality class:
This Feynman diagram can be evaluated by iteratively applying our method.
The diagram gives
\begin{align}
    \int dx_1dx_2 \, G(p,x,x_1) G'(p,x_1,x_4)G(p,x_4,0)\,,
\end{align}
where $G'(p,x_1,x_4)$ can be written as,
\begin{align}
    \label{eq:gp}
    G'(p,x_1,x_4)=\int d^{d-1}y\,e^{-ip y}(-i\gamma^5)G''(y,x_1,x_4)(-i\gamma^5)D(y,x_1,x_4)\,.
\end{align}
The function $G''$ has a similar structure. 
Therefore, we have 
\begin{align}
    \label{eq:gpp}
    G''(y,x_1,x_4)=\int \frac{d^{d-1}k}{(2\pi)^{d-1}}dx_2dx_3\,e^{ik y}G(k,x_1,x_2)G'''(k,x_2,x_3)G(k,x_3,x_4)\,, 
\end{align}
where $G'(p,x_1,x_4)$ can be written as,
\begin{align}
    \label{eq:gppp}
    G'''(k,x_3,x_4)=\int d^{d-1}y' \,e^{-ik y'}(-i\gamma^5)G(y',x_2,x_3)(-i\gamma^5)D(y',x_2,x_3)\,.
\end{align}
In the presence of a boundary, the fermion propagator can be decomposed into two components, $G = G_s + G_b$, where $G_b$ and $G_s$ represent the parts with and without translation symmetry, respectively. 
Similar to the one-loop calculation explicitly carried out in our work, the Fourier transformation in Eq.~\eqref{eq:gppp} contains four terms, $G_bD_b$, $G_sD_s$, $G_bD_s$ and $G_sD_b$.  
Plugging the terms after Fourier transformation into Eq.~\eqref{eq:gpp}, and performing the integration over real coordinates $x_2$ and $x_3$, one obtains the function depending on $|x_1-x_4|$, $(x_1+x_4)$ and the momentum $k$. 
Bring the momentum $k$ back to the real coordinate $y$, one thus gets $G''(y,x_1, x_4)$. Finally, a similar procedure can be carried out starting from Eq.~\eqref{eq:gp}, and the evaluation of this Feynman diagram is straightforward.

\section{Renormalization group calculation for the Chiral XY universality class}
In this section, we calculate the boundary critical exponent for the chiral XY universality class. 
The chiral XY universality class can arise at the quantum phase transition between a Dirac semimetal and a Kekul\'e valence bond solid on a honeycomb lattice when the flavor of the Dirac fermion is large enough~\cite{li2017fermion,jian2017fermion}. 
The GNY action for the chiral XY universality class reads,
\begin{align}
    S=\int_{\mathcal M} d^d x\Bigg( \sum_j\bar\psi_j\gamma^\mu\partial_\mu\psi_j+\frac{1}{2}\sum_{i=1,2}(\partial\phi_i)^2+\frac{\lambda}{4!}\Big(\sum_{i=1,2}\phi_i^2\Big)^2+g\sum_{j}\Big(\phi_1\bar\psi_j\psi_j+i\phi_2\bar\psi_j\gamma^5\psi_j\Big)\Bigg).
\end{align}
In the absence of boundaries, the one-loop RG factors are well known~\cite{zerf2017four-loop}:
\begin{align}
    &Z_\phi=1-\frac{Ng^2}{4\pi^2\epsilon},\qquad Z_\psi=1-\frac{g^2}{8\pi^2\epsilon}, \\
    &Z_g=1+\frac{(1+N)g^2}{8\pi^2\epsilon},\qquad Z_\lambda=1+\frac{Ng^2}{2\pi^2\epsilon}+\frac{5\lambda}{24\pi^2\epsilon}-\frac{3Ng^4}{\lambda\pi^2\epsilon}.
\end{align}
Hence, the RG equation reads
\begin{align}
    \frac{{\rm d}\lambda}{{\rm d\ log}\mu}&=-\epsilon\lambda+\frac{Ng^2\lambda}{2\pi^2}+\frac{5\lambda^2}{24\pi^2}-\frac{3Ng^4}{\pi^2},\\
    \frac{{\rm d}g}{{\rm d\ log}\mu}&=-\frac{\epsilon}{2}g+\frac{g^3}{8\pi^2}+\frac{Ng^3}{8\pi^2}.
\end{align}
It leads to the chiral XY fixed point at
\begin{align}
    \lambda^*=\frac{12\pi^2(1-N+\sqrt{1+38N+N^2})}{5(1+N)}\epsilon, \quad
    g^*=\frac{2\pi^2\sqrt{\epsilon}}{\sqrt{1+N}}.
\end{align}
Consequently, the scaling dimensions for the boson and fermion are, respectively, given by
\begin{align}
    \Delta_{\phi}=1-\frac{1}{2+2N}\epsilon, \quad 
    \Delta_{\psi}=\frac{3}{2}-\frac{1+2N}{4+4N}\epsilon.
\end{align}

Let's move on to the boundary criticality and take the Kekul\'e valence bond solid transition as an example. 
To be concrete, we consider a ribbon with two armchair boundaries. 
Recall that the two-component order parameter $\phi_{1,2}$ behaves as a ``$\mathbb Z_3$'' vector~\cite{li2017fermion} under translation transformation. 
Now, because the translation symmetry is explicitly broken by the open boundaries, the boundary mass terms for $\phi_1$ and $\phi_2$ are no longer related.
Moreover, it turns out that in the ribbon geometry, there is a mirror symmetry $x \rightarrow -x$ that can forbid the linear term for $\phi_2$, but in general, a linear term in $\phi_1$ is allowed on the boundary. 
Hence, the boundary boson $\phi_1$ will obey a normal boundary condition due to the presence of a linear term, while the boundary boson $\phi_2$ can still satisfy distinct boundary conditions determined by the mass term. 
The normal boundary condition in the GNY universality class is still an outstanding question, and we leave the investigation to a future work. 
In the following, we assume that the linear term in $\phi_1$ is absent, and consider independent boundary conditions for the two components $\phi_{1,2}$, respectively. 

Due to the lack of symmetry relation on the boundary, the boson boundary conditions can be classified into four cases:
\begin{enumerate}
    \item Both \(\phi_1\) and \(\phi_2\) satisfy the Neumann boundary condition.
    \item Both \(\phi_1\) and \(\phi_2\) satisfy the Dirichlet boundary condition.
    \item \(\phi_1\) satisfies the Neumann boundary condition while \(\phi_2\) satisfies the Dirichlet boundary condition.
    \item \(\phi_1\) satisfies the Dirichlet boundary condition while \(\phi_2\) satisfies the Neumann boundary condition.
\end{enumerate}

In most cases, the calculations follow the same procedures as for the chiral Ising universality class. 
However, due to the lack of a symmetry operation that forbids the $\phi_1$ component, as detailed in the previous paragraph and in the main text,
it gives rise to an additional Feynman diagram, shown in Fig.~2(d), to the boundary fermion correction. 
This Feynman diagram leads to the following contribution to the correlation function $\langle \psi \hat {\bar \psi} \rangle$,
\begin{align} \label{eq:G_N}
    & \int dx_1dx_2G(k,x,x_1)\mathbb{I}G(k,x_1,0)D(k',x_1,x_2) \left(-{\rm Tr}\left[\mathbb{I}G(y=0,x_2,x_2)\right] \right), \\
    =& \int dx_1dx_2G(k,x,x_1)G(k,x_1,0)D(k',x_1,x_1)\times\left(-\frac{x_2^{-3+\epsilon}}{4\pi^2\epsilon}\right)\nonumber\\=&\frac{1}{8\pi^2\epsilon}\left[\frac{i\slashed k}{2q}-\frac{\gamma^1}{2}-\gamma^1\left(\frac{i\slashed k}{2q}+\frac{\gamma^1}{2}\right)\right]e^{-qx}.
\end{align}
Incorporating this difference, the final results at the one-loop order are summarized below. 
\begin{enumerate}
    \item Both $\phi_1$ and $\phi_2$ satisfy the Neumann boundary condition:
    \begin{align*}
        \Delta_{\hat\psi}&=\frac32-\frac{1+2N}{2+2N}\epsilon, \\
        \Delta_{\hat\phi_1}&=1-\frac{6+9N+\sqrt{1+38N+N^2}}{10+10N}\epsilon, \\
        \Delta_{\hat\phi_2}&=1-\frac{6-N+\sqrt{1+38N+N^2}}{10+10N}\epsilon, \\
        \Delta_{\hat\phi_1^2}&=2-\frac{8+7N-2\sqrt{1+38N+N^2}}{10+10N}\epsilon, \\
        \Delta_{\hat\phi_2^2}&=2-\frac{8-3N-2\sqrt{1+38N+N^2}}{10+10N}\epsilon.
    \end{align*}
    
    \item Both $\phi_1$ and $\phi_2$ satisfy the Dirichlet boundary condition:
    \bea
        \Delta_{\hat\psi} & =& \frac32-\frac{2N+1}{2+2N} \epsilon, \\
        \Delta_{\partial\hat\phi_1}&=&2-\frac{6-N+\sqrt{1+38N+N^2}}{10+10N}\epsilon, \\
        \Delta_{\partial\hat\phi_2}&=&2-\frac{6+9N+\sqrt{1+38N+N^2}}{10+10N}\epsilon.
    \eea
    \item $\phi_1$ satisfies the Neumann boundary condition while $\phi_2$ satisfies the Dirichlet boundary condition:
    \begin{align*}
        \Delta_{\hat\psi}&=\frac32-\frac{3+4N}{4+4N}\epsilon, \\
        \Delta_{\hat\phi_1}&=1-\frac{6+9N+\sqrt{1+38N+N^2}}{10+10N}\epsilon,\\
        \Delta_{\hat\phi_1^2}&=2-\frac{8+7N-2\sqrt{1+38N+N^2}}{10+10N}\epsilon,\\
        \Delta_{\partial\hat\phi_2}&=2-\frac{6+9N+\sqrt{1+38N+N^2}}{10+10N}\epsilon.
    \end{align*}
    \item $\phi_1$ satisfies the Dirichlet boundary condition while $\phi_2$ satisfies the Neumann boundary condition:
    \begin{align*}
        \Delta_{\hat\psi}&=\frac32-\frac{1+4N}{4+4N}\epsilon, \\
        \Delta_{\partial\hat\phi_1}&=2-\frac{6-N+\sqrt{1+38N+N^2}}{10+10N}\epsilon, \\
        \Delta_{\hat\phi_2}&=1-\frac{6-N+\sqrt{1+38N+N^2}}{10+10N}\epsilon,\\
        \Delta_{\hat{\phi}_2^2}&=2-\frac{8-3N-2\sqrt{1+38N+N^2}}{10+10N}\epsilon.
    \end{align*}
\end{enumerate}

\end{document}